\catcode`\@=11

\newif\if@fewtab\@fewtabtrue

{\count255=\time\divide\count255 by 60 \xdef\hourmin{\number\count255}
\multiply\count255 by-60\advance\count255 by\time
\xdef\hourmin{\hourmin:\ifnum\count255<10 0\fi\the\count255}}
\def\ps@draft{\let\@mkboth\@gobbletwo
    \def\@oddhead{}
    \def\@oddfoot
       {\hbox to 7 cm{$\scriptstyle Draft\ version:\ \draftdate$
       \hfil}\hskip -7cm\hfil\rm\thepage \hfil}
    \def\@evenhead{}\let\@evenfoot\@oddfoot}

\def\ceqno{\global\@fewtabfalse
    \ifcase\@eqcnt \def\@tempa{& & &}\or \def\@tempa{& &}
      \or \def\@tempa{&}
      \or\def\@tempa{}\fi\@tempa
{\rm(\theequation)}}

\def\aeqno#1{\global\@fewtabfalse
    \ifcase\@eqcnt \def\@tempa{& & &}\or \def\@tempa{& &}
      \or \def\@tempa{&}
      \or\def\@tempa{}\fi\@tempa
{\rm(\theequation,#1)}}

\def\label#1{\ifnum\draftcontrol=1
 \global\def\draftnote{$\scriptstyle #1$}\fi
 \@bsphack\if@filesw {\let\thepage\relax
   \def\protect{\noexpand\noexpand\noexpand}%
\xdef\@gtempa{\write\@auxout{\string
      \newlabel{#1}{{\@currentlabel}{\thepage}}}}}\@gtempa
   \if@nobreak \ifvmode\nobreak\fi\fi\fi
  \@esphack}

\def\alabel#1#2{\label{#1}\global\@fewtabfalse
    \ifcase\@eqcnt \def\@tempa{& & &}\or \def\@tempa{& &}
      \or \def\@tempa{&}
      \or\def\@tempa{}\fi\@tempa
{\hbox to 3cm{\phantom{\rm(\theequation,#2)}
\draftnote \hfil}\hskip -3cm {\rm(\theequation,#2)}}}

\def\clabel#1{\label{#1}\global\@fewtabfalse
    \ifcase\@eqcnt \def\@tempa{& & &}\or \def\@tempa{& &}
      \or \def\@tempa{&}
      \or\def\@tempa{}\fi\@tempa
{\hbox to 3cm{\phantom{\rm(\theequation)}
\draftnote \hfil}\hskip -3cm{\rm(\theequation)}}}

\def\eqnarray{\def\draftnote{{}}\global\@fewtabtrue
\stepcounter{equation}\let\@currentlabel=\theequation
\global\@eqnswtrue
\global\@eqcnt\z@\tabskip\@centering\let\\=\@eqncr
$$\halign to \displaywidth\bgroup\@eqnsel\hskip\@centering
  $\displaystyle\tabskip\z@{##}$&\global\@eqcnt\@ne
  \hskip 2\arraycolsep \hfil${##}$\hfil
  &\global\@eqcnt\tw@ \hskip 2\arraycolsep $\displaystyle\tabskip\z@{##}$
\hfil  \tabskip\@centering&\global\@eqcnt\thr@@\llap{##}\tabskip\z@\cr}

\def\endeqnarray{\@@eqncr\egroup
      \global\advance\c@equation\m@ne$$\global\@ignoretrue}

\def\@eqnnum{\hbox to 3cm{\phantom{\rm(\theequation)} \draftnote
                         \hfil}\hskip -3cm {\rm(\theequation)}}

\def\@@eqncr{\let\@tempa\relax
    \ifcase\@eqcnt \def\@tempa{& & &}\or \def\@tempa{& &}
      \or \def\@tempa{&}
      \or\@fewtabfalse
\fi
\if@eqnsw\if@fewtab\@tempa
\@eqnnum\fi\stepcounter{equation}\fi\global
\@eqnswtrue\global\@eqcnt\z@\global\@fewtabtrue\cr}

\def\sime{\mathop{\triangleright\!\!\!<}}

\font\tendl=msym10  scaled \magstep1
\font\sevendl=msym7 scaled \magstep1
\font\fivedl=msym5 scaled \magstep1
\font\tengl=eufm10  scaled \magstep1
\font\sevengl=eufm7 scaled \magstep1
\font\fivegl=eufm5 scaled \magstep1

\newfam\dlfam  
\textfont\dlfam=\tendl \scriptfont\dlfam=\sevendl
\scriptscriptfont\dlfam=\fivedl
\newfam\glfam  
\textfont\glfam=\tengl \scriptfont\glfam=\sevengl
\scriptscriptfont\glfam=\fivegl

\def\draftdate{\number\month/\number\day/\number\year\ \ \ \hourmin }

\global\def\draftcontrol{0}
\catcode`\@=12
\def\tilde{\widetilde}
\documentstyle[12pt]{article}
\newcommand{\be}{\begin{eqnarray}}
\newcommand{\en}{\end{eqnarray}\vs 0.5 cm}

\newcommand{\no}{\noindent}
\newcommand{\vs}{\vskip}
\newcommand{\hs}{\hspace}

\newcommand{\p}{\partial}
\newcommand{\un}{\underline}

\newcommand{\NR}{{{\bf R}}}
\newcommand{\NC}{{{\bf C}}}
\newcommand{\NZ}{{{\bf Z}}}
\newcommand{\qq}{\begin{eqnarray}}

\newcommand{\ee}{{\rm e}}

\newcommand{\qqq}{\end{eqnarray}\vskip-0.1cm}

\newcommand{\tr}{\hbox{\rm tr}}

\newcommand{\ra}{\rightarrow}

\newcommand{\B}{{\rm B}}

\newcommand{\CD}{{\cal D}}

\newcommand{\CF}{{\cal F}}
\newcommand{\CG}{{\cal G}}
\newcommand{\CH}{{\cal H}}

\newcommand{\CK}{{\cal K}}

\newcommand{\CO}{{\cal O}}

\newcommand{\CQ}{{\cal Q}}

\newcommand{\CU}{{\cal U}}
\newcommand{\CV}{{\cal V}}

\newcommand{\J}{{\rm J}}
\newcommand{\s}{{\hspace{.05cm}}}
\newcommand{\half}{{_1\over^2}}
\newcommand{\h}{{h^{\hspace{-0.24cm}-}}}
\begin{document}
\
\vskip 1.5cm



\date{ }
\begin{center}
{\large\bf{LATTICE WESS-ZUMINO-WITTEN MODEL
\vskip 0.3cm

AND QUANTUM GROUPS}}\footnote{Extended
version of lectures given by the second author at
the XXVIII\s th Karpacz Winter School
of Theoretical Physics, February 17 - 29, 1992}
\vskip 1.3cm
Fernando FALCETO\footnote{Present address: Department of Theoretical Physics,
University of Zaragoza, Zaragoza, Spain}

I.H.E.S., 91440 Bures-sur-Yvette, France
\vskip 0.2cm

Krzysztof GAW\c{E}DZKI

C.N.R.S., I.H.E.S., 91440 Bures-sur Yvette, France,
\end{center}
\vskip 1.5cm


\begin{abstract}
Quantum groups play a role of symmetries of
integrable theories in two dimensions. They may be detected
on the classical level as Poisson-Lie
symmetries of the
corresponding phase spaces. We discuss specifically
the Wess-Zumino-Witten conformally invariant
quantum field model combining two chiral parts
which describe the left- and right-moving degrees of freedom.
On one hand side, the quantum group plays the role
of the symmetry of the chiral components of the theory.
On the other hand, the model admits a lattice regularization
(in the Minkowski space) in which the current algebra symmetry
of the theory also becomes quantum, providing
the simplest example of a quantum group symmetry coupling space-time
and internal degrees of freedom. We develop a free
field approach to the representation
theory of the lattice $sl(2)$-based current algebra and show how to use
it to rigorously construct an exact solution of the
quantum $SL(2)$ WZW model on lattice.
\end{abstract}
\vs 1.5 cm

\no{\bf 1.\ \ Introduction}
\vs 0.4cm

The relations to two-dimensional quantum field theories and
integrable lattice models have
been at the origin of quantum groups \cite{1},\cite{2}, objects abstracted
from the remarkable properties of exactly soluble two-dimensional
systems. The language of quantum groups, as that of the standard groups,
has become a convenient tool to describe symmetry
properties of physical systems. Quantum groups are one-parameter
deformations of classical Lie groups or, if we identify the deformation
parameter with $\h$, their quantization. A more
rewarding point of view is however the one in which, on the classical
level, we retain the information about the infinitesimal direction
of the deformation. We obtain then the notion of a Poisson-Lie (PL)
group, i.e. a Lie group with a compatible Poisson
bracket on it \cite{3},\cite{4}.
\vs 0.2cm

The standard symmetries of classical systems are described by
the action of the symmetry group elements on the phase space of the system
in a way which preserves
the Poisson brackets of physical quantities. The infinitesimal
version of the action of continuous symmetry groups on physical
quantities may then be generated by Poisson bracket with hamiltonians
whose collection goes under the name of a moment map \cite{5}.
Giving the moment map is an alternative way to describe
the classical symmetry.
\vs 0.2cm

PL groups lead to a generalization of the concept of
classical symmetry. One requires that not the individual
symmetry transformations but the action of the group viewed as
a transformation from \s$phase\ space\times group\s$ into \s$phase\ space\s$
preserves Poisson brackets. The notion of a moment map may be
generalized to the case of PL symmetries: the infinitesimal
transformations of physical quantities are still generated by
Poisson brackets, this time with a non-linear collection of
hamiltonians (in terminology advocated in \cite{6}).
\vs 0.2cm

When quantizing a classical system with standard symmetries, one often
attempts to preserve the symmetries, i.e. to get the action of the
same symmetry groups in the quantum space of states in such a way
that the interesting families of observables transform in a simple way
mimicking their classical covariance properties (this is not always
possible). When quantizing a classical system with a PL
symmetry, it is natural to demand an action of the corresponding
quantum group in the space of states. The covariance properties
of the interesting observables should then reduce to the classical
expressions in order $\CO(\h)$. The theory of PL symmetries
becomes this way very useful as the means to detect quantum group
symmetries of quantum mechanical or quantum field-theoretical models.
\vs 0.2cm

The present set of lectures is designed as an illustrative
introduction to the subject of Poisson-Lie and quantum group
symmetries. We start by a presentation of the  elements
of the theory of PL groups (Section 2) which may be found
in numerous original and review papers, see e.g. \cite{1},
\cite{3},\cite{4},\cite{37},
\cite{7},\cite{8},\cite{6}.
The main body of the exposition discusses an application
of those concepts to the Wess-Zumino-Witten (WZW) model of
conformal field theory. In Section 3 we describe
the canonical formalism of the classical WZW theory
following the approach of \cite{9}, see also \cite{37}-\cite{13}.
A special stress is put on the details of how the degrees of freedom
of the model separate into the left- and right-movers.
We show that, besides the standard loop group (or current algebra)
and conformal (Virasoro) symmetries, the chiral part of the
classical theory posses a finite-dimensional PL symmetry.
In Section 4, we demonstrate how the last two symmetries may be
separated by decomposing the chiral phase space of the
WZW model. This decomposition is the classical counterpart
of the vertex-IRF transformation well known from the theory
of lattice integrable systems \cite{14} or from the analysis
of the quantum WZW theory \cite{15}. The loop group part
of the chiral phase space is essentially a union
of coadjoint orbits of the infinite-dimensional Kac-Moody group
centrally extending the loop group. Similarly, the Poisson-Lie part
is composed of PL deformations of the coadjoint orbits
for the finite dimensional PL group. Under quantization,
they give spaces, respectively, of irreducible representations of
Kac-Moody algebra and (Drinfeld-Jimbo \cite{1},\cite{2}) quantum group.
The space
of quantum states of the chiral WZW theory combines both. In Section 5,
we describe a free field realization of the loop group part
of the chiral WZW phase space for the simplest case
of group $SL(2)$, see also \cite{12}. The free fields provide Darboux
coordinates
on the phase space and a convenient starting point for
the Wakimoto-Bernard-Felder approach to the representation
theory of $sl(2)$-based  Kac-Moody algebra. This approach realizes
the irreducible
representations of the latter in a cohomology of Fock spaces.
\vs 0.2cm

As most quantum theories, the WZW model requires a (field strength)
renormalization.
This complicates the rigorous construction of the theory
which involves fine and interesting mathematical points far from being
completely clarified  (most members of the conformal
physics community are probably not aware of this state of affairs).
It is then interesting to realize that the model possesses
a lattice regularization which essentially preserves the continuum
symmetries of the theories. For the related Liouville and Toda theories
such regularizations where discussed in \cite{38} to \cite{42}.
For the WZW theory, the lattice regularization may be based
on the concept of the lattice Kac-Moody algebra introduced
in a series of papers of the Leningrad-St.\s Petersburg
group \cite{16}-\cite{18}.
The regularization may be introduced already on the
classical level in such a way that the theory still separates
into the left- and right-moving sectors and that
the loop group symmetry of
the continuum theory becomes a local Poisson-Lie symmetry,
as described in Section 6. The classical vertex-IRF transformation
permits to separate further the degrees of freedom carrying
the local and global PL symmetries. In the quantum theory,
the moment map of the local PL symmetry becomes the St. Petersburg
lattice Kac-Moody algebra. We present it (for the simplest $sl(2)$ case)
in Section 7.
One of its remarkable features is that it implies a lattice
version of conformal symmetry formulated as an invariance
under a block spin type renormalization group! In fact the lattice
Kac-Moody algebra realizes a simplest coupling of the
quantum groups to the space-time degrees of freedom
and might be thought of as an elementary step towards employing
quantum groups as space-time symmetries. The quantum space
of states corresponding to the lattice Kac-Moody algebra
may be built from special representations of the latter
obtained by a deformation of the free field approach
{\it \`{a} la} Wakimoto-Bernard-Felder. This is described in
Sections 8 to 10 containing an original material, see also \cite{19}.
In the simplest case of one lattice
site, the lattice Kac-Moody algebra reduces to the
(Drinfeld-Jimbo) quantum group. As a result, by specialization,
the free field construction of representations of lattice
Kac-Moody algebra provides also a possible approach to
the representation theory of the quantum group. We provide
its details in Section 11 which may be read independently
and/or prior to Sections 8 to 10. Finally, in Section 12,
we combine the spaces of states of lattice Kac-Moody
and quantum group degrees of freedom and show how dressing
of the first ones with the second allows to obtain
field operators with braiding given by the quantum $SL(2)$
$R$-matrices in fundamental representation. This provides
a lattice realization of the idea spelled out in \cite{15}.
\vs 0.2cm

As discussed in \cite{20}-\cite{22}, the continuum chiral
WZW model is closely
related to the three-dimensional topological Chern-Simons gauge theory:
it essentially describes the Schr\"{o}dinger picture of the latter.
The main feature of the representation theory of the
Kac-Moody algebras which underlies the relation between the
WZW and the Chern-Simons models is the existence of the special
tensor product of representations of Kac-Moody algebras
called fusion. There are many indications that there should
exist a lattice regularization of the Chern-Simons model, a type
of a quantum-group gauge theory, which is still topological
(invariant under lattice subdivisions).
In fact, the expectation
values of the Chern-Simons theory (giving 3-manifold
and knot invariants) may be computed as
lattice statistical sums involving quantum group objects
\cite{23},\cite{24}.
We expect that the chiral lattice
WZW theory described in these lectures may be a possible
step towards a lattice Chern-Simons topological
quantum group gauge theory. The main missing step seems
to be the lattice fusion operation which, up to now,
we have under complete control only if one of the fused
representations corresponds to the lattice with a single point.
\vs 0.2cm

Another open problem, equally interesting, is to find
a lattice version of the coset construction which,
in continuum, allows to produce a multitude of
conformal field theories out from the WZW one. In particular,
the lattice version of Drinfeld-Sokolov reduction should
reproduce the lattice versions \cite{40} to \cite{42} of Toda
theories.
\vs 1cm

\no{\bf 2.\ \ Poisson-Lie symmetry}
\vs 0.3cm
\no{\bf 2.a.\ \ Poisson manifolds}
\vs 0.4cm

One of the essential concepts of the classical physics is that of the Poisson
bracket of physical quantities. Mathematically, one considers manifolds $M$
with a (smooth) field of 2-vectors $\s\Pi
\s( = \sum\limits_{i,j} \Pi^{ij}\, {\p
\over \p x^{i}} \wedge {\p \over \p x^j} $ in local coordinates\s)
\s s.~t.~the Poisson bracket of functions $f,g$ on $M$ given by the contraction
of
$\Pi$ with 2-form \s$df \wedge dg$\s:
$$\{ f,g \} = \langle \Pi, df \wedge dg \rangle \eqno (1)$$
satisfies the Jacobi identity.
$$\{ \{ f,g \}, h \} + \{ \{ g,h\},f \} + \{\{ h,f \},g \} = 0\ .\eqno (2)$$
$\Pi$ is called then a Poisson structure on $M$. $M$ together with the Poisson
bracket is called a Poisson manifold.
There are two extreme cases of Poisson structures: $\Pi \equiv 0$ and $\Pi$
non-degenerate, i.e. $\det(\Pi^{ij}) = 0$. In the latter case, where
necessarily
$M$ is even-dimensional, $\Pi$ induces a symplectic form $\Omega$ on $M$ equal
locally to $\sum\limits_{i,j}{}(\prod^{-1})_{ij} dx^{j} \wedge dx^i$. On
symplectic
manifolds, the only functions which Poisson commute with all others are
constants. In general, $M$ may be foliated into (maximal connected) manifolds,
called symplectic leaves, on which each function on $M$ Poisson commuting with
the others is constant and on which the Poisson structure induces a symplectic
form. If \s$\Pi \equiv 0$\s, \s the symplectic leaves are the points of \s$M$.
\vs 0.2cm

For two Poisson manifolds \s$(M_i, \Pi_i),\ i = 1,2$\s,  $\s\Pi_1 + \Pi_2\s$
defines a Poisson structure on $M_1 \times M_2$. \s If $\s f_i,\s g_i\s$
are functions on $\s M_i\s$, respectively, then
$$\{ f_1 f_2\s,\s g_1g_2 \} = \{ f_1\s,\s g_1\} f_2g_2 + f_1g_1
\{f_2\s,\s g_2 \}\ .\eqno (3)$$
A smooth map \s$P : M_1 \ra M_2\s$ is called Poisson if on $P(M_1)$
$$P_* \Pi_1 = \Pi_2 \eqno (4)$$
where $P_*$ denotes the tangent map. Then, for functions \s$f,\s g\s$ on $M_2$,
$$\{ P^*f, P^*g \} = P^* \{f,g \}\ . \eqno (5)$$
\vs 0.2cm

\no{\bf 2.b.\ \ Poisson-Lie groups and Lie bialgebras}
\vs 0.3cm
Crossbreeding of the concepts of a Poisson manifold and of a Lie group has
led to the notion of a Poisson-Lie (PL) group \cite{3},\cite{4}.
This is a Lie group supplied
with a Poisson structure such that the group multiplication map
$$ m : G \times G \mapsto G \eqno (6) $$
is Poisson. Below, we shall work in the complex
category, unless otherwise stated. On the infinitesimal level, one
obtains from the Poisson
structure of $G$ the Lie algebra bracket $[\cdot,\cdot]^*$
on the dual space $\CG^*$
to the Lie algebra $\CG$ of $G$\s:
$$[df(1),dg(1)]^* = d \{ f,g\}(1)\ . \eqno (7)$$
The fact that the group multiplication is a Poisson map leads to the
following compatibility condition between the Lie brackets of $\CG$ and
$\CG^*$:
$$\langle [X,Y],[v,w]^* \rangle + \langle ad_v^*X,ad_Y^*w \rangle - \langle
ad_w^* X, ad_Y^* v \rangle - \langle ad_v^*Y, ad_X^* w \rangle + \langle
ad_w^* Y, ad_X^* v \rangle = 0\ . \eqno (8) $$
By dualization, $[ \cdot , \cdot ]^*$ induces a map
$$c : \CG \ra \CG \wedge \CG\ .$$
In terms of \s$c\s$, \s eq.~(8) becomes the cocycle
condition $$c[X,Y] - ad_X \, c(Y) + ad_Y\, c(X) = 0\ . \eqno (9) $$
A pair of finite-dimensional Lie algebras $(\CG , \CG^*)$ with Lie brackets
satisfying (8) is called a Lie bialgebra. Each Lie bialgebra corresponds to a
unique connected, simply connected PL group. Because $\CG$ and $\CG^*$ enter
symmetrically into (8), $(\CG^*,\CG)$ is
also a Lie bialgebra which we shall call
dual to $(\CG,\CG^*)$. The PL group $(G^*, \{ \cdot,\cdot \}^*)$ corresponding
to $(\CG^*,\CG)$ will be called dual to $(G, \{ \cdot , \cdot \} )$. Thus
(connected, simply connected) PL groups come in dual pairs.
\vs 0.3cm

$\CG\s$ and \s$\CG^*\s$ may be put as Lie subalgebras into the
double Lie algebra \s$\tilde\CG\s$, \s equal \s$\CG\oplus\CG^*\s$
as the vector space, with the Lie bracket
$$[X+v\s,\s Y+w]\equiv [X,Y]+[v,w]^*-ad_X^*w+ad_Y^*v+ad_w^*X-ad_v^*Y\s.$$
\no They  form maximal isotropic subspaces of \s$\tilde\CG\s$ with
respect to $ad$-invariant non-degenerate bilinear form
$$(X+v\s,\s Y+w)\equiv\langle X,w\rangle+\langle Y,v\rangle\s.$$
The (simply connected) group $\tilde G$
corresponding to \s$\tilde\CG\s$ is the
Drinfeld double of PL group $(G,\{.,.\})$.
\vs 0.2cm

Conversely, any Lie algebra \s$\tilde\CG\s$ with a non-degenerate
$ad$-invariant bilinear form and a pair of maximal isotropic
subalgebras (a Manin triple) gives (a pair of dual) Lie bialgebras
by identifying one of the subalgebras with the dual of the other
by means of the bilinear form.
\vs 0.2cm
\eject

\no{\bf 2.c.\ \ Examples of PL groups}
\vs 0.3cm

Each Lie group may be considered as a PL group if we take $\{ \cdot, \cdot \}
\equiv 0\s$. \s The corresponding Lie bialgebra has trivial bracket
$\s[ \cdot , \cdot ]^*
\equiv 0\s$ on $\s\CG^*\s$. The dual PL group is $\CG^*$ itself viewed as the
additive group together with the Poisson bracket
$$ \{ X,Y \}^* = [X,Y] \eqno (10)$$
where $X,Y,[X,Y] \in \CG$ are viewed as linear functions on $\CG^*$. The
symplectic leaves of $\CG^*$ are the (connected components of) coadjoint orbits
of $G$. They are models of symplectic homogeneous spaces of $G$ and
they give rise, by quantization, to irreducible representations
of $G$ \cite{25}.
\vs 0.3cm

An (almost) general example of a PL group is
provided by the following construction.
Let $r = \sum\limits_{\alpha}{} X_\alpha
\otimes Y_\alpha
\in \wedge^2 \CG$. Define a bracket on $\CG^*$ by
$$ \langle X , [v,w]^* \rangle =
\langle ad_X r\s,\s v \otimes w \rangle \ . \eqno
(11)$$
The corresponding cocycle $c : \CG \ra \CG \wedge \CG$ is given by
$$c(X) = ad_X r \eqno (12) $$
i.e.~is a coboundary. Bracket $[\s \cdot , \cdot ]^*$ satisfies the Jacobi
identity if and only if
$$ ad_X \, r_{123} = 0 \eqno (13)$$
for each $X \in \CG$ where
$$r_{123} =  [r_{12}, r_{13}] + [r_{12},r_{23}] + [r_{13},r_{23}]J\in
\CG^{\wedge 3} \eqno (14)$$
and
$$[r_{12},r_{13}] = \sum\limits_{\alpha,\beta}{} [X_\alpha,X_\beta]  \otimes
Y_\alpha
\otimes Y_\beta \in \CG^{\otimes 3}\s,\ \ \ {\rm etc.}$$
\vs 0.2cm

Suppose that $\CG$ is a (complex) simple
Lie algebra, with the Killing form $\s\tr\s$ and a basis $\s(t^{a})\s$
normalized so that $\s\tr\s\s t^{a}t^b = \half
\delta^{ab}$. Any ad-invariant element of $\CG^{\wedge 3}$ is proportional to
$\s F =
\sum\limits_{a,b,c}{} f^{abc}\, t^{a} \otimes t^b \otimes t^c\s$ \s where
\s$[t^{a}, t^b] =
\sum\limits_{c}{}f^{abc} t^c\s$, \s so that \s$r_{123} \sim F\s$ or,
equivalently,
denoting
the proportionality constant as $\s- ( {_{2 \pi} \over^k})^2\s$ for later
convenience,  $$r^\pm_{123} = 0 \eqno (15)$$ where
$$r^{\pm} \equiv r \pm {_{2 \pi} \over^ k} C \eqno (16) $$
with the quadratic Casimir \s$C = \sum\limits_{a}{}\, t^{a} \otimes t^{a}\s$.
Condition (15) is known as the classical Yang-Baxter (CYB) equation.
A possible choice for $r$ is
$$ r = {_\pi \over^k}\sum\limits_{\alpha > 0}{} (e_{\alpha} \otimes e_{-
\alpha}
- e_{- \alpha}
\otimes e_{\alpha}) \eqno (17)$$
where $e_\alpha$ are the step generators
of $\CG$ corresponding to roots $\alpha$. We
shall refer to the corresponding $r^\pm$ as the standard solutions of the CYB
equation. In the case of simple $\CG$ (and ${_{2 \pi} \over^k} \not = 0$) one
may
realize $\CG^*$ with its Lie bracket (11) as a subalgebra of $\CG \oplus \CG$
by the embedding
$$ \CG^* \ni v \mapsto \bigl (\s\iota(v) r^+,\s\iota (v)r^- \bigr ) \in \CG
\oplus
\CG \eqno (18)$$
which is a Lie algebra homomorphism due to the CYB equation (14).
$\iota(v)r^\pm$
denotes the contraction of $v$ with the first component of $r^\pm$. For the
standard $r$-matrix (17), one obtains this way the subgroup of
\s$b^- \oplus b^+\s$
\s(\s$b^\pm$ are the Borel subgroups
of $\CG$\s) \s composed of elements with opposite Cartan subalgebra components.
\vs 0.2cm

The Poisson bracket of the PL group $\s(G,\{.,.\})\equiv G_r\s$
corresponding to
Lie bialgebra $(\CG,\CG^*)$ with bracket (11) may
be described the following way.
Let $g$ denote the matrix of functions on $G$ given by a finite-dimensional
representation of $G$. Let \s$g_1 = g \otimes 1,\ g_2 = 1 \otimes g\s$ where
$1$
is the unit matrix. Then the formula
$$\{ g_1\s,\s g_2 \} = [\s g_1g_2\s,\s r\s]
\s\s\s(\s= [\s g_1g_2\s,\s r^\pm\s]\s) \eqno (19)$$
where $r$ is taken in the same representation, gives the Poisson bracket of
all matrix elements of $g$.
For simple $\CG$, the dual PL group \s$G_r^*\s$ may be realized as a
subgroup of \s$G\times G\s$. \s For the standard $r$-matrix one
obtains the subgroup of \s$B^- \times
B^+\s$ \s(\s$B^\pm\s$ are the Borel subgroups) composed of elements
\s$(\gamma^-, \gamma^+) =
(n^-t,n^+t^{-1})\s$ \s where $\s n^\pm\s$ are in the nilpotent subgroups
and $t$ is in the Cartan subgroup $\s T\s$ of \s$G\s$.
The Poisson bracket $\{ \cdot , \cdot \}^*$ on \s$G^*\s$ is given by the
formulae
$$\matrix {
\{ \gamma^+_1, \gamma^+_2 \}^*   & = & [\s r\s, \gamma_1^+ \gamma_2^+ ]\ ,  \cr
\cr
\{ \gamma_1^+, \gamma_2^- \}^*   & = & [\s r^+\s,\gamma_1^+ \gamma_2^- ]\ ,
\cr
\cr
\{ \gamma_1^-, \gamma_2^- \}^*   & = & [\s r\s, \gamma_1^-  \gamma_2^- ]\ .
\cr
} \eqno (20)
$$
By the map $\s(\gamma^-,\gamma^+) \s\smash{\mathop{\mapsto}\limits_{}^{\chi}}\s
 \gamma^-(\gamma^+)^{-1}=\gamma\s,$\ $\s G^*\s$
covers finitely many times an open subset of $G$ (which, for the standard
$r$-matrix is also dense). The Poisson bracket $\{ \cdot, \cdot \}^*$
$$\{ \gamma_1,\gamma_2 \}^* = r^+\gamma_1 \gamma_2 - \gamma_1\s r^+
\gamma_2 - \gamma_2\s r^- \gamma_1
+ \gamma_1 \gamma_2\s r^- \eqno (21)$$
makes the covering map Poisson. (The Poisson bracket $\{ \cdot , \cdot \}^*$,
as contrasted with the one of (19) does not make $G$ a PL group). The conjugacy
classes in $G$ form the symplectic leaves of $(G, \{ \cdot , \cdot \}^*)$. The
connected components of the $\chi$-preimages of the conjugacy classes form the
symplectic leaves of $G^*\s$ \cite{4} \s which play in the PL
category the role generalizing
that of the coadjoint orbits in the Lie category. In particular,
for standard $r$-matrix, their quantization gives the irreducible
representations of quantum group $\s\CU_q(\CG)\s$ deforming
the universal enveloping algebra of $\CG$.
\vs 0.4cm

\no{\bf 2.d.\ \ Moment maps}
\vs 0.3cm

Let \s$(M,\{ \cdot , \cdot \})\s$ be a Poisson manifold.
We say that Lie group $G$ is a
symmetry of $M$ if $G$ acts on $M$ (say, from the right) by
transformations  preserving the Poisson
bracket. If $X \in \CG$, it generates then a vector field $\tilde X$ on $M$.
We shall call $\tilde X$ hamiltonian if there exists a function $h_X$ s.~t.
$$\tilde X(f) = \{ h_X,f \} \eqno (22)$$
for any smooth function $f$ on $M$. If $M$ is symplectic and simply connected
then $h_X$ always exists. If $\tilde X$ and $\tilde Y$ are hamiltonian
then so is $\widetilde{aX + bY}$ and $[ \widetilde{X,Y}]$ and we may take
$$h_{aX + bY} = a\s h_X + b\s h_Y\s, \eqno (23)$$
$$h_{[X,Y]} = \{ h_X, h_Y \}\ . \eqno (24)$$
The action of $G$ is called hamiltonian if each $\tilde X$ is hamiltonian. The
hamiltonians $h_X$ may be chosen so that (23) is satisfied. Then we obtain a
map
$$m : M \ra \CG^*$$
s.~t. $\s\langle X,m (\cdot) \rangle = h_X\s$. \s If additionally
(24) is satisfied, \s$m\s$
is a Poisson map if we provide $\CG^*$ with Poisson bracket (10). In this case
$\s m\s$ is called the moment map for the action of $G$ on $M$ \cite{5}.
If $\CG$ is simple,
(24) may be always assured. In general, one has to pass to a central extension
of $G$ and $\CG$.
\vs 0.2cm

The notion of a symmetry may be naturally generalized to the PL category. We
shall say that a PL group $\s(G,\{ \cdot , \cdot \})\s$ is a PL symmetry of the
Poisson manifold $\s(M, \{ \cdot , \cdot \})\s$
if the map \s$M \times G \ra M\s$ giving
the action of \s$G\s$ on \s$M\s$ is Poisson.
If the Poisson bracket on \s$G\s$ is trivial,
this reduces to the previous notion of the symmetry.
\s We shall call a PL symmetry hamiltonian if there exists a map
$$m : M \ra G^* $$
s.t.
$$\tilde X(f) = \langle X, \{ m,f \} m^{-1} \rangle \eqno (25)$$
for each smooth function $f$ on $M$ \s\s(\s
$\{ m,f \}m^{-1} \equiv \langle\s \Pi\s,\s (dm)m^{-1}\wedge df\s
\rangle\s$ where \s$\Pi\s$
is the Poisson structure on \s$M$\s). \s
If \s$M\s$
is symplectic and simply connected, then each PL symmetry is hamiltonian
\cite{8}. We
shall say that \s$m\s$ is a moment map if,
additionally, it is a Poisson map. Its
existence is not guaranteed even for
hamiltonian symmetry. In general, \s$m\s$ is
Poisson only for a modified Poisson structure of $G^*$ \cite{8}.
Clearly, the moment map allows to reconstruct the (infinitesimal) action of $G$
on $M$.
\vs 1cm

\no {\bf 3.\ \ Classical Wess-Zumino-Witten theory}
\vs 0.4cm

The Wess-Zumino-Witten (WZW) theory is an example of a two-dimensional
sigma model with fields $g(x^0,x^1)$ taking values in the group manifold $G$.
The classical equations are
$$ \p_-( g \p_+ g^{-1}) = 0 \eqno (1)$$
where $\p_{\pm}$ are the derivatives with respect to the light-cone
variables
$$x^{\pm} = x^1 \pm x^0\ . \eqno (2)$$
We shall consider the theory on the cylinder obtained by taking $x^1$ modulo $2
\pi$. The general solutions of (1) are
$$g(x^0,x^1) = g_L(x^+)\s g_R(x^-)^{-1} \eqno (3) $$
where
$$ g_{L,R} (x^\pm + 2 \pi) = g_{L,R}(x^\pm)\s\gamma \eqno (4)$$
with monodromy $\s\gamma \in G.\ \s(g_L,g_R)\s$ and $\s(g_Lg,g_Rg)\s$
with constant $\s g\s$
give the same solution. If we introduce spaces \s$P_{L,R} = \big\{ g_{L,R} :
\NR\ra G \mid g_{L,R} (x+ 2 \pi)
= g_{L,R}(x) \gamma_{L,R} \s\big\}$\s, \s then the phase
space \s$P\s$ of solutions of (1) becomes
$$P = \Delta/G \eqno (5)$$
where $\Delta$ is the submanifold of $P_L \times P_R$ where $\gamma_L =
\gamma_R$.
\s$P_{L,R}$ will play the role of phase spaces for the left-, right-moving
degrees
of freedom.
\vs 0.3cm

Eq. (1) comes from the Lagrangian
\addtocounter{equation}{5}
\renewcommand{\theequation}{\arabic{equation}}
\qq
S(g) = {_k \over^{ 4 \pi}} \int {\rm tr}\s(g^{-1} \p_+g)\s(g^{-1}
\p_- g)\s dx^+ \wedge dx^-
+ {_k \over^{ 12 \pi}} \int d^{-1}{\rm  tr}\s (g^{-1}dg)^{\wedge 3}
\qqq
\no ($k$ is the coupling constant) which
induces the canonical Poisson bracket on
$P\, $:
\qq
\matrix{\hbox to 3.8cm{$\{g(0,x)_1\s,\s g(0,y) \}$\hfill}\ =\  0\ \ \ \
,\hfill\cr
\hbox to 3.8cm{$\{ g(0,x)_1\s,\s \xi^0(0,y)_2 \}$\hfill}\ =\
-{{8 \pi}\over k} \s g(0,x)_1 \, C \s\delta(x-y)\ ,\hfill\cr
\hbox to 3.8cm{$\{\xi ^0 (0,x)_1\s,\s \xi^0(0,y)_2 \}$\hfill}\ =\
{{8 \pi}\over k}\s \left[\s C\s,\s
\xi^0(0,x)_1 + g^{-1} \p_1g(0,x)_1 \s \right ]\s \delta(x-y)}
\qqq

\no where $\xi^0 \equiv g^{-1} \p_0g$.
The corresponding symplecting form is
$$ \Omega = {_k \over^{4 \pi}} \int_{0}^{2 \pi} \tr \s\left [ - d \xi^0 \wedge
(g^{-1} dg) + (\xi^0 + g^{-1} \p_1g)(g^{-1} dg)^{\wedge 2} \right ] (0,x)\ .
\eqno (8)$$
Rewriting \s$\Omega\s$ in terms of $g_{L,R}$, we obtain
$$\Omega(g) = \Omega_{L}(g_L) - \Omega_R(g_R)\ \ \ \ \  \eqno (9)$$
where
\addtocounter{equation}{2}
\renewcommand{\theequation}{\arabic{equation}}
\qq
\Omega_L(g_L) = {_k \over^{4 \pi}} \int_{0}^{2 \pi} \s{\rm tr\s} (g^{-1}_Ldg_L)
\wedge \p (g_L^{-1}dg_L)\cr
+ {_k \over^{ 4 \pi}} \s{\rm tr\s}
 g_L^{-1} dg_L (0) \wedge (d \gamma_L) \gamma_L^{-1} - {_k\over^{ 4
\pi}} \rho(\gamma_L)
\qqq
\no and \s$\Omega_R\s$ is given by the same formula with $g_R \ra g_L$. $\rho$
is an
arbitrary 2-form on $G$. The ambiguity is due to the fact that $\Omega$ is
given
only on the subset $\Delta$ of $P_L \times P_R$. Since
$$d \Omega_L(g_L) = {_k \over^{         12 \pi}} {\s\rm tr\s}
(\gamma_L^{-1} d \gamma_L)^{\wedge 3} - {_k \over^{4
\pi}} d \rho (\gamma_L)\hspace{0.8cm}\eqno (11)$$
and \s$\tr\s(\gamma^{-1}d \gamma)^{\wedge 3}\s$ is a closed but
not exact form on (simple) $G$,
\s$\Omega_L$ is never globally closed. One may however choose singular
\s$\rho\s$
s.~t. \s$d
\rho = {_1 \over^3} \tr\s
(\gamma^{-1} d \gamma)^{\wedge 3}\s$ on an open (dense) set. It will
be convenient to take
$$\rho(\gamma) = \tr\s\s
(\gamma^-)^{-1} (d \gamma^-) \wedge (\gamma^+)^{-1}(d \gamma^+)\hspace{0.6cm}
\eqno (12)$$
in terms of the decomposition $\gamma = \gamma^-(\gamma^+)^{-1}$ induced by a
classical
$r$-matrix. This choice leads to the following, globally defined, Poisson
bracket on \s$P_L\s$:
\vs -0.4cm

\addtocounter{equation}{2}
\renewcommand{\theequation}{\arabic{equation}}
\qq
\hbox to 3cm{$\{ g_L(x)_1\s,\s g_L(y)_2 \}$\hfill}
&=& g_L(x)_1 \, g_L(y)_2\,  r^\pm \ ,\\
\hbox to 3cm{$\{ g_L(x)_1\s,\s \gamma_{L2} \}$\hfill}
&=& g_L(x)_1 \gamma_{L2} \s
r^- - g_L(x)_1 \, r^{+} \gamma_{L2} \ ,\\
\hbox to 3cm{$\{ \gamma_{L1}\s,\s \gamma_{L2} \}$\hfill}
&=& r^+\gamma_{L1},
\gamma_{L2} - \gamma_{L1}\,  r^{+} \gamma_{L2} -
\gamma_{L2} \s r^{-}\gamma_{L1} + {\gamma}_{L1} \gamma_{L2}\s r^-
\qqq
\no where in (13) the upper sign in
$r^\pm$ applies when \s$x > y\s$ and the lower one
when \s$x < y\s$, $|x-y|<2\pi\s$.
\s The left-moving phase space $P_L$
equipped with the Poisson bracket (13) - (15)
possesses a rich symmetry structure.
\vs 0.3cm

\no{\un{Conformal symmetry}}
\vs 0.2cm

Let \s$\widetilde {\rm Diff}_+ S^1\s$
denote the group of increasing maps \s$D : \NR \ra
\NR\s$ s.~t. \s$D(x+ 2 \pi) = D(x) + 2 \pi\s.$ \s
$\widetilde {\rm Diff}_+ S^1$ is a central
extension by its subgroup $\NZ$ of integer shifts of group ${\rm Diff}_+ S^1$
of
orientation preserving diffeomorphisms of $S^1$ :
$$0 \ra \NZ \ra \widetilde{\rm Diff}_+ S^1 \ra {\rm Diff}_+ S^1 \ra 1\ .$$
$\widetilde {\rm Diff}_+ S^1 $ acts on $P_L$ by
$$g_L \mapsto g_L \circ D$$
preserving the Poisson brackets (13) - (15). The moment map is given by the
energy momentum tensor
$$T_L(x) = {_k \over^{4 \pi}} \tr\s(g_L^{-1} \p_x g_L)^2 \, (dx)^2 \eqno (16)$$
defining, as a quadratic differential, an element dual to ${\rm Vect}(S^1)$,
i.e. to the vector fields on the circle $\NR/(2 \pi \NZ)$ which form the Lie
algebra
of $\widetilde {\rm Diff}_+ S^1$. \s The \s$\widetilde {\rm Diff}_+ S^1\s$
symmetry of \s$P_L\s$
(and another copy of it for \s$P_R$\s) \s give rise to the action on
\s$P\s$ of the group
\s$(\widetilde{\rm Diff}_+ S^1 \times \widetilde{\rm Diff}_+ S^1)/\NZ\s$
 of conformal
symmetries of the cylinder (the quotient subgroup is that of the diagonal
shifts).
\vs 0.3cm

\no{\un{Loop group symmetry}}
\vs 0.2cm

Another symmetry of $P_L$ is given by the
action of the loop group $LG$ of smooth maps $h : \NR \ra G,\ h(x+2 \pi) =
h(x)$\s,
\s by  $$g_L \mapsto h^{-1} g_L\ .$$
This action is hamiltonian and the currents
$$J(x) = {_k \over^{2 \pi}} \s g_L \p g_L^{-1} dx \eqno (17) $$
give the corresponding map of $P_L$ into the space $L\CG{}^*$ dual to the Lie
algebra $L \CG$ of $LG$. This map is, however, not Poisson due to the $\delta'$
term
in the Poisson bracket
$$\{ J(x)_1\s,\s J(y)_2 \} = 2\s [\s C\s,\s J(x)_1 ]\,\s  \delta(x-y) -
{_k \over^\pi}\,  C\s\delta'(x-y)\ .\eqno (18)$$
We may however pass to the central extension of $L\CG$
$$0 \ra \NC \ra \hat{L\CG} \ra L\CG \ra 0$$
equal $L\CG \oplus \NC$ as a vector space where $\NC$ is central and
$$[X,Y](x) = [X(x), Y(x)] +  {_i \over^{2 \pi}} \int_{0}^{2 \pi}\s \tr\s X'Y$$
for $X,Y \in  L\CG \subset \hat{L\CG}$. \s$\hat{L\CG}$ is the Kac-Moody
algebra. On a global level, one obtains the Kac-Moody group $\hat{LG}$
giving the central extension of $LG$
$$1 \ra \NC^* \ra \hat{LG} \ra LG \ra 1\ .$$
$J +{_1\over^i} k \in L{\CG}{}^* \oplus \NC \cong \hat{L\CG}^*$ defines
then the moment
map for the $\hat{LG}$ symmetry of $P_L$ (with the center of $\hat{ LG}$
acting trivially). On the complete phase space we obtain the $\hat {LG} \times
\hat {LG}$ symmetry.
\vs 0.4cm

\no{\underline{$G_r\s$ Poisson-Lie symmetry}}
\vs 0.3cm

The conformal and loop-group symmetries of $P_L$ are the standard ones: the
action of group elements preserves the Poisson bracket on $P_L$. \s$P_L$
possesses also a Poisson-Lie symmetry given by the action of $G$
$$(g_L,g) \mapsto g_Lg$$
which becomes a Poisson map if we consider $G$ as a $PL$ group
equipped with Poisson
bracket (2.19). The monodromy map
$$g_L \mapsto \gamma$$
plays essentially the role of the moment map for this action. More exactly, its
local lifts to $G^* \subset G \times G$
$$g_L \mapsto (\gamma^-,\gamma^+)$$
s.~t. $\gamma = \gamma^-(\gamma^+)^{-1}$ satisfy the following relations
$$\matrix{
\{ \gamma_1^+,\s g_L(x)_2 \} &= &-g_L(x)_2 \,  r^+ \gamma_1^+\ , \cr
\{ \gamma^-_1,\s g_L(x)_2 \} &= &-g_L(x)_2\,  r^- \gamma_1^- \ , \cr
\{ \gamma^+_1, \gamma_2 \}\hfill &= &[\s r^+,\gamma_2\s]\s\gamma^+_1\ , \hfill
\cr
\{\gamma_1^-,\s\gamma_2 \}\hfill &= &[\s r^-,\gamma_2\s]\s\gamma_1^- \ .\hfill
\cr
}$$
These are equivalent to the defining relation (2.25) of the moment map.
\vs 1cm

\no {\bf 4.\ \ Classical vertex-IRF transformation}
\vs 0.4cm

On the open dense subset $P_{L0}$ of $P_L$ composed of $g_L$ with monodromies
$\gamma$ in the conjugacy classes of elements of the Cartan subgroup $T \subset
G$, we
may use the following parametrization
$$g_L(x) = h(x)\s\ee^{i \tau x} g_0^{-1} \eqno (1) $$
where $h \in LG, \ \tau \in t$ (the Cartan algebra) and $g_0 \in G$.
Notice that $$\gamma=g_0\s\ee^{2\pi i\tau}\s g_0^{-1}\ .\eqno (2)$$
Parametrization (1) defines a map from $LG \times t \times G_0$ onto $P_{L_0}$.
Let $Q^\vee \subset t$ denote the coroot lattice and $N(T)$ the normalizer of
\s$T$
in $G$. We shall assume $G$ simply connected. A semi-direct product $N(T) \sime
Q^\vee$ acts on $LG \times t \times G_0$ by
$$\bigl (h(x),\tau,g_0 \bigr)\s \mapsto \s\bigl (h(x)\s \ee^{iqx} w\s,\s
w^{-1} (\tau -
q)w\s,\s g_0w \bigr )\eqno (3)$$ for \s$w \in N(T)\s,\ \s q \in Q^\vee\s$.
\s Its orbits
describe exactly the ambiguity of parametrization (1). In terms of (1), the
2-form $\Omega_L$ of (3.10) becomes
$$\Omega_L(g_L) = \Omega_{L1}(h,\tau) + \Omega_{L2}(g_0,\tau) \eqno (4)$$
where
$$\Omega_{L1}(h,\tau) = {_k \over^{4 \pi}} \int_{0}^{2 \pi} \tr\s \left
[(h^{-1} dh)
\wedge \p (h^{-1}dh) + 2 i \tau(h^{-1}dh)^{\wedge 2} - 2 i\, d\, \tau \wedge
h^{-1} dh \right ] \eqno (5)$$
and
$$
\Omega_{L2} (g_0,\tau)
=
 {_k \over^{4 \pi}}\s\tr\s (g_0^{-1} dg_0) \wedge \ee^{2
\pi i \tau} (g^{-1}_0 dg_0)\s \ee^{- 2 \pi i \tau}
+ ki\s\tr\s d \tau \wedge g_0^{-1} dg_0 -{_k \over^{4 \pi}}\s
\rho (g_0\s\ee^{2 \pi i \tau} g_0^{-1} )\ .
\eqno (6) $$
$\Omega_{L1}\s$ and \s$\Omega_{L2}\s$ are closed forms
(we could also subtract from $\Omega_{L1}$
an arbitrary closed 2-form dependent on $\tau$ and add it to $\Omega_{L2}$).
Space
$M_{\hat{LG}} \equiv LG \times t$ with
symplectic form $\Omega_{L1}$ is (a covering) of what
is called a model space for the Kac-Moody group. When restricted to fixed $\tau
\in t$, $\Omega_{L1}$ coincides with the pullback to $LG$ of the canonical
symplectic
form on the coadjoint orbit
$$\left \{ h(i \tau)h^{-1} + h \p h^{-1} +
{_1\over^i}k  \mid h \in LG \right\} \subset
\hat{L \CG}^*\ .$$
$\tau, \tau'$ correspond to the same coadjoint orbit if and only if $\tau' =
w(\tau + q) w^{-1}$ for $q \in Q^\vee$ and $w \in N(T)$\s, \s i.e.
when $\tau$ and
$\tau'$ are related by the affine Weyl group. Thus the coadjoint orbits appear
in \s$M_{\hat{LG}}\s$ with multiplicity given by
the affine Weyl group. Similarly, \s$M_{G_r}
\equiv G \times t\s$ plays the role of the (covering) of the model space for
the
PL group $G_r$ with Poisson bracket (2.19). For fixed $\tau \in t,\
\Omega_{L2}$ coincides with the pullback to $G$ of the symplectic form on the
conjugacy class
$$\{ g_0 \s\ee^{2 \pi i \tau} g_0^{-1} \mid g_0 \in G \} $$
which is a symplectic leaf of the Poisson structure (2.21). We remind that the
latter corresponds to that of the dual $PL$ group $G_r^*$ under the map
\s$(\gamma^-,\gamma^+) \mapsto \gamma^-(\gamma^+)^{-1}\s$.
\s Again, $\tau$ and $\tau'$ give
the same conjugacy class if and only if
they are related by the affine Weyl group.
Let \s$\delta \subset M_{\hat{LG}} \times M_{G_r}\s$ denote the
submanifold given by equating the \s$t\s$
coordinates. Then \s$P_{L0} = \delta/N(T) \sime Q^\vee\s$.
\s As we see, \s$P_{L0}\s$ may be
foliated by leaves being products of coadjoint orbits of $\hat{LG}$ and
conjugacy classes of $G$ corresponding to the same $\tau$, each such pair
appearing only once.
\vs 0.3cm

\addtocounter{equation}{-10}
\renewcommand{\theequation}{\arabic{equation}}
For the compact real form of group $G$, model space $M_{\hat{LG}}$
may be quantized geometrically. The resulting
space of states is (if we translate the original $k$
to $k+h^{\hspace{-0.05cm}\check{}}$ where $h^{\hspace{-0.05cm}\check{}}$
is the dual Coxeter number and use the half-form geometric quantization)
\be
\CH_{\hat{LG}}\ =\ \bigoplus\limits_{{\rm integrable}\ \lambda}
\CH_{k,\lambda}\s.
\en
\vs -0.4cm
\no $\CH_{k,\lambda}\s$ is the space of the irreducible unitary
representation of \s$\hat{LG}\s$ of level \s$k\s$ and
highest weight \s$\lambda\s$
coming from geometric quantization of the coadjoint orbit corresponding
to $\tau=(k+h^{\hspace{-0.05cm}\check{}}\hspace{0.03cm})^{-1}(\lambda+\rho)$.
$\rho\equiv\sum\limits_{\alpha>0}\alpha/2$ is the Weyl vector.
Fields \s$h_L(x)\s$ become the vertex operators (in the
fundamental representation) \cite{26}.
\vs 0.2cm

One may attempt a quantization of model space $M_{G_r}$ along
similar lines obtaining space
\qq
\CV=\bigoplus\limits_{{\rm integrable}\ \lambda}\CV_{q,\lambda}
\qqq
\no where $\s\CV_{q,\lambda}\s$ is the space of irreducible representation
of quantum group $\s\CU_q(\CG)\s$ for
\s$q=\ee^{\pi i/(k+h^{\hspace{-0.05cm}\check{}}\hspace{0.03cm})}\s$
of highest weight \s$\lambda\s$
coming from the quantization of the symplectic leaf
of $G_r^*$ corresponding to the conjugacy class of $\s\ee^{2\pi i
(k+h^{\hspace{-0.05cm}\check{}}\hspace{0.03cm})^{-1}
(\lambda+\rho)}$\s. Quantum fields
$g_0$ become the Clebsch-Gordan coefficients of $\s\CU_q(\CG)\s$
(quantum $3j$-symbols).
\s The space of states
of the chiral WZW model quantizing the phase space $P_L$ would then be
given by the sum of diagonal products
$$\bigoplus\limits_{{\rm integrable}\ \lambda}\CH_{k,\lambda}\otimes
\CV_{q,\lambda}\s.$$
Below, we shall realize the quantization program for the lattice
regularization of the WZW model using free field representation of
the theory rather than geometric quantization.
\vs 1 cm

\no{ \bf 5.\ \ Free field realization of the WZW model}
\vskip 0.4cm
\addtocounter{equation}{-7}
\renewcommand{\theequation}{\arabic{equation}}

Model space $M_{\hat{LG}}$ may be realized in terms of free fields which
provide Darboux coordinates for it. Let us describe the construction
for the case $G=SL(2,\NC)$. Parametrize:
\be
h_L(x)=\pmatrix{1&v(x)\cr 0&1}\pmatrix{1&0\cr w(x)&1}
\pmatrix{\ee^{i\phi(x)}&0\cr 0&\ee^{-i\phi(x)}}
\en
\no where $v$ and $w$ are periodic and
$$\phi(x+2\pi)=\phi(x)+2\pi p$$
($\tau=p\s\sigma_3$). Put
\be
\beta=v\s,\ \ \ \ \ \ \gamma=\ee^{-2i\phi}\s
\p(\ee^{2i\phi}w)=\p w+2i(\p\phi)w\s.
\en
\no $w$ may be recovered from $\phi$ and $\gamma$:
$$w(x)=(\ee^{4\pi ip}-1)^{-1}\s\ee^{-2i\phi(x)}\int
\limits_x^{x+2\pi}\gamma(y)\s\ee^{2i\phi(y)}\s dy\s.$$
\no In terms of $\phi,\ \beta$ and $\gamma$.
\be
\Omega_{L1}(h,\tau)={_k\over^{2\pi}}\int d\beta\wedge d
\gamma-{_k\over^{2\pi}}\int d\phi\wedge\p(d\phi)
+k\s d p\wedge d\phi(0)\s,
\en
\vs -0.2cm
\no so that the nontrivial Poisson brackets are
\be
\{\phi(x)\s,\s\phi(y)\}&=&{_\pi\over^{2k}}\s{\rm sgn}(y-x)\s,\cr
\{\phi(x)\s,\s p\}\ \ \ \ &=&{_1\over^{2k}}\s,\cr
\{\gamma(x)\s,\s\beta(y)\}&=&{_{2\pi}\over^k}\s\delta(x-y)\s.
\en
\vs -0.5cm
\no For the fields, we obtain
\be
h_L(x)\s=\pmatrix{(\Pi-\Pi^{-1})\s\Psi(x)+
\beta(x)\s\Psi(x)^{-1}\s Q(x)&\ \beta(x)\s\Psi(x)^{-1}\cr
\Psi(x)^{-1}\s Q(x)&\ \Psi(x)^{-1}}\pmatrix{(\Pi-\Pi^{-1})^{-1}&0\cr 0&1}
\label{vert}
\en
\vs -0.4cm
\no where \ $\Pi\equiv\ee^{2\pi ip}\s,\ \ \Psi(x)\equiv\ee^{i\phi(x)}$
\ and the integral of the screening charge
\qq
Q(x)\equiv\Pi^{-1}\int\limits_x^{2\pi+x}\gamma(y)\s\Psi(y)^2\s dy\s.
\label{SC}
\qqq
\vs -0.2cm
\no Current
\be
J={_k\over^{2\pi}}\s h_L\s\p h_L^{-1}\s dx=
{_k\over^{2\pi}}\pmatrix{-i\p\phi-\beta\gamma&\ -\p\beta+2i(\p\phi)\beta
+\beta^2\gamma\cr \gamma&\ i\p\phi+\beta\gamma}dx\s.\label{cur}
\en
\vs 0.3 cm

As we see, $M_{\hat{LG}}$ may be described via classical scalar field $\phi$
and classical $\beta\gamma$-system (essentially a complex scalar field).
Both may be quantized in Fock spaces. The quantum version of (\ref{cur})
may be produced by Wick ordering the products of free fields.
One obtains this way Fock space representations of $\hat{L\s sl}(2)$
Kac-Moody algebra (the so called Wakimoto modules) \cite{27}.
The irreducible unitary representations of
$\hat{L\s sl}(2)$ may be realized in the cohomology of a complex of Fock space
ones, with the differential obtained from various (regularized) powers
of the integral $Q$ of the screening operator \cite{28}. The quantum version
of field $\s h_L(x)\s$ may again be constructed by Wick-ordering
(\ref{vert}). For the details, we refer to the original work \cite{28}.
A lattice
version of this construction will be presented in Sections 7 to 9.
\vs 1cm

\no{\bf 6.\ \ WZW theory on lattice}
\vskip 0.4cm
\addtocounter{equation}{-7}
\renewcommand{\theequation}{\arabic{equation}}

Let us consider the lattice cylinder $\NZ\times\NZ$ with identification
$(n^+,n^-)\cong(n^++N,n^-+N)$ ($n^+,n^-$ are integer
values of the light-cone variables
$x^\pm$). The following difference equation is the lattice counterpart
of classical equation of motion (2.1) of the WZW theory:
\be
g(n^+,n^-)\s\s g(n^++1,n^-)^{-1}=g(n^+,n^--1)\s\s g(n^++1,n^--1)^{-1}\s.
\en
\vs -0.3cm
\no The space of solutions may be parametrized by the values $g(n,n)$
and $g(n+1,n)$ of the lattice field $g$ at two subsequent time
slices (the lattice Cauchy data). The general solution has the form
\be
g(n^+,n^-)=g_L(n^+)\s\s g_R(n^-)^{-1}
\en
\vs -0.4cm
\no where
\vs -0.4cm
\be
g_{L,R}(n+N)=g_{L,R}(n)\s\s\gamma_{L,R}\s.\label{ph-s}
\en
\vs -0.2cm
\no with $\gamma_L=\gamma_R$. \ Let $P_{L,R}^N$ denote the space
of $g_{L,R}$ satisfying (\ref{ph-s}). Again, space of
all solutions of (1) is
$$P^N=\Delta/G$$
where $\Delta$ is the submanifold of $P^N_L\times
P^N_R$ given by the
equation $\gamma_L=\gamma_R$. We shall consider $P^N_L$ with the
Poisson bracket given by
\be
\hbox to 3.15cm{$\{g_L(n)_1\s,\s g_L(m)_2\}$\hfill}
&=&g_L(n)_1\s g_L(m)_2\s\s r^\pm\s,\cr
\hbox to 3.15cm{$\{g_L(n)_1\s,\s g_L(n)_2\}$\hfill}
&=&r^+\s g_L(n)_1\s g_L(n)_2+g_L(n)_1\s
g_L(n)_2\s\s r^-\s,\cr
\hbox to 3.15cm{$
\{g_L(n)_1\s,\s \gamma_{L2}\}$\hfill}
&=&g_L(n)_1\s\gamma_{L2}\s\s r^--
g_L(n)_1\s\s r^+\s\gamma_{L2}\s,\cr
\hbox to 3.15cm{$\{\gamma_{L1}\s,\s\gamma_{L2}\}$\hfill}
&=&r^+\s\gamma_{L1}\s\gamma_{L2}\s
-\gamma_{L1}\s\s r^+\s\gamma_{L2}\s-\gamma_{L2}\s\s r^-\s\gamma_{L2}
+\s\gamma_{L1}\s\gamma_{L2}\s\s r^-\s.\label{ChPB}
\qqq
\no where in the first line $\pm$ corresponds to $n{_>\atop^<}m$\s,
\ $|n-m|<N$. $\ P^N_R$ has the Poisson
bracket of opposite sign. These two brackets
induce a (local) Poisson bracket on $P^N$ which for the Cauchy
data takes the form:
\be
\hbox to 5cm{$\{g(n,n)_1\s,\s g(n,n)_2\}$\hfill}&=&
[\s r\s,\s g(n,n)_1\s g(n,n)_2\s]\s,\cr
\hbox to 5cm{$\{g(n,n)_1\s,\s g(m,m)_2\}$\hfill}&=&
0\ \ \ \ {\rm for}\ \ m\not=n\s\s{\rm mod}\s N\s,\cr
\hbox to 5cm{$\{g(n+1,n)_1\s,\s g(n+1,n)_2\}$\hfill}&=&
[\s r\s,\s g(n+1,n)_1\s g(n+1,n)_2\s]\s,\cr
\hbox to 5cm{$\{g(n+1,n)_1\s,\s g(m+1,m)_2\}$\hfill}&=&
0\ \ \ \ {\rm for}\ \ m\not=n\s\s{\rm mod}\s N\s,\cr
\hbox to 5cm{$\{g(n,n)_1\s,\s g(n+1,n)_2\}$\hfill}&=&
-g(n,n)_1\s g(n+1,n)_2\s\s r^+\s,\cr
\hbox to 5cm{$\{g(n,n)_1\s,\s g(n,n-1)_2\}$\hfill}&=&
r^-\s\s g(n,n)_1\s g(n,n-1)_2\s,\cr
\hbox to 5cm{$\{g(n,n)_1\s,\s g(m+1,m)_2\}$\hfill}&=&
0\ \ \ \ {\rm for}\ \ m\not=n,\s n-1\s\s{\rm mod}\s N\s.
\en
\vs 0.4cm

\no\underline{Lattice conformal symmetry}
\vs 0.3cm

The above Poisson brackets on $P^N_{L,R}$ and $P^N$
are invariant under shifts
\be
g_{L,R}(\cdot)\s\ \smash{\mathop{\longmapsto}
\limits^{\CU_{L,R}}}\s\ g_{L,R}(\cdot+1)\s\s.
\en
\no In general, if $D:\NZ\rightarrow \NZ$ is an increasing map
s.~t. \s$D(n+N')=D(n)+N\s$ (necessarily, $N'\leq N)$ then it induces
a Poisson map
\be
P^N_{L,R}\ni g_{L,R}(\cdot)\s\ \smash{\mathop{\longmapsto}
\limits^{\CD_{L,R}}}\s\ g_{L,R}(D(\cdot))\in P^{N'}_{L,R}\s\s.
\en
\no These maps, of the form of decimation, constitute the lattice versions
of the conformal transformations: on the lattice the conformal invariance
is expressed as invariance under a renormalization group type
transformations.
\vs 0.4cm

\no\underline{Lattice loop group symmetry}
\vs 0.3cm

$G^N$ acts on $P^N_L$ by
\be
g_L(n)\s\mapsto\s h(n)^{-1}\s g_L(n)\s.
\en
\no This action does not preserve the Poisson structure of $P^N_L$
but it becomes a Poisson-Lie symmetry if we consider each component
of $G^N$ as a PL group $G_r$ with Poisson bracket (2.19).
Let us recall that in the continuum, current $J=(k/2\pi)\s g_L\s\p g_L^{-1}$
was almost the moment map for the loop group symmetry of $P_L$.
Its Poisson brackets corresponded to the canonical bracket on
the dual $\hat{L\CG}^*$ of the central extension of the loop group
Lie algebra. Similarly, the lattice version of the current
\be
J(n)=g(n+1)g(n)^{-1}\label{curr}
\en
\no does not correspond to the Poisson bracket on $(G_r^*)^N$ as
should the moment map for PL symmetry $G_r^N$ but to its deformation:
\be
\hbox to 3.4cm{$\{J(n)_1\s,\s J(n)_2\}$\hfill}&=&
r^+\s J(n)_1\s J(n)_2+J(n)_1\s J(n)_2\s\s r^-\s,\cr
\hbox to 3.4cm{$\{J(n)_1\s,\s J(n+1)_2\}$\hfill}&=&
-J(n+1)_2\s\s r^+\s J(n)_1\s,\clabel{latcurr}\cr
\hbox to 3.4cm{$\{J(n)_1\s,\s J(m)_2\}$\hfill}&=&
0\ \ \ \ {\rm if}\ m\not=n-1,n,n+1\s\s{\rm mod}\s N\s.
\en
\vs 0.2cm
\no$G^N$ with the Poisson bracket (\ref{latcurr})
describes the semiclassical object
which corresponds to the lattice Kac-Moody algebra which was
introduced in \cite{16},\cite{17} and which we shall denote $\CK_N$ below.
We shall quantize the lattice
WZW theory basing on the representation theory of \s$\CK_N\s$.
\vs 0.4cm

\no\underline{$G_r$ Poisson-Lie symmetry}
\vs 0.3cm

Exactly as in the continuum case, the map
$$P^N_L\times G_r\ni(g_L,g)\s\mapsto\s g_Lg\in P^N_L$$
gives a PL symmetry of $P^N_L$ whose moment map is (locally)
given by the lift of the monodromy $\s\gamma\in G\s$ to $\s(\gamma^+,
\gamma^-)\in G_r^*$\s.
\vs 0.5cm

Again as in continuum, we may parametrize
\qq
g_L(n)=h(n)\s\ee^{in\tau}\s g_0^{-1}\equiv h_L(n)\s g_0^{-1}
\qqq
\no where $\s(h,\tau)\s$ are coordinates of the model space $M_{\CK_N}$
of the (semiclassical version of) lattice Kac-Moody algebra $\CK_N$
and $\s(g_0,\tau)\s$ of the model space $M_{G_r}$ introduced above.
The space of quantum states corresponding to $M_{\CK_N}$ will be
a discrete sum of representations of $\CK_N$ and that corresponding to
$P_L^N$ will be a sum of their combinations with irreducible representations
of quantum group $\CU_q(\CG)$. Quantum $\s h_L(n)\s$ will be built
of vertex operators of lattice Kac-Moody algebra $\CK_N$ and quantum
chiral WZW fields $\s g_L(n)\s$ will combine $\s h_L(n)\s$ with
quantum $\s g_0^{-1}\s$ composed of $\CU_q(\CG)$ vertex (or Clebsch-Gordan)
operators. The rest of the present exposition will be devoted to
the realization of this program for $G=SL(2)$.

\vs 1cm

\no{ \bf 7.\ \ Lattice Kac-Moody algebra}
\vs 0.4cm
\addtocounter{equation}{-11}
\renewcommand{\theequation}{\arabic{equation}}

The solutions $r^\pm$ of CYB equations may be quantized in representations,
at least in some cases, to the solutions $R^\pm$ of the quantum version of
the Yang-Baxter equation (without spectral parameter):
\be
R^\pm_{12}\s R^\pm_{13}\s R^\pm_{23}=R^\pm_{23}\s R^\pm_{13}\s R^\pm_{12}\s.
\en
\no$R^+=P\s (R^-)^{-1} P$\s where $\s P\s$\s interchanges the factors in the
tensor product and \s$R^\pm=1+{_1\over^i}r^\pm+\CO(k^{-2})\s.$
\s This may be done for example for the standard $r$-matrix (2.17).
For $SL(2)$ with which, for simplicity, we shall work from now on,
\be
R^+\ =\ q^{1/2}\s\pmatrix{q^{-1}&0&0&0\cr 0&1&q^{-1}-q&0\cr
0&0&1&0\cr 0&0&0&q^{-1}}\ ,\hs{1cm}R^-\ =\ q^{-1/2}\s\pmatrix{q&0&0&0\cr
0&1&0&0\cr 0&q-q^{-1}&1&0\cr 0&0&0&q}\label{R}
\qqq
\no in the fundamental representation.
\s$q=\ee^{\pi i/(k+2)}\s$. \s$SL(2)$-based lattice Kac-Moody
algebra $\CK_N$ \cite{16},\cite{17}
is an associative algebra with generators organized into $2\times 2$
matrices ($n$ is taken modulo $N$)
\be
\J(n)\s\equiv\s\pmatrix{\J(n)_{11}&\J(n)_{12}\cr
\J(n)_{21}&\J(n)_{22}}\s\in\s\CK_N\otimes{\rm End}(\NC^2)\s\s,
\qqq
\vs -0.2cm
\no such that
\vs -0.2cm
\qq
\J(n)_{11}\s \J(n)_{22}-q^{-1}\s \J(n)_{21}\s
\J(n)_{12}=q^{1/2}
\qqq
\vs -0.2cm
\no and with relations
\vs -0.2cm
\qq
\hbox to 2.8 cm{$\J(n)_1\s \J(n)_2\s$\hfill}&=&
\s R^{+}\s \J(n)_2\s \J(n)_1\s\s R^{-}\s,\clabel{LKM}\cr
\hbox to 2.8cm{$\J(n)_1\s \J(n+1)_2\s$\hfill}&=&
\s \J(n+1)_2\s\s(R^+)^{-1}\s \J(n)_1\s,\cr
\hbox to 2.8cm{$\J(n)_1\s \J(m)_2$\hfill}&=&
\J(m)_2\s \J(n)_1\ \ \ \ {\rm for}\ m\not=n-1,n,n+1\s.
\qqq
\no Notice that the semi-classical versions of
those relations
obtained by expanding to the leading order in $k^{-1}$ and replacing
the commutator by $_1\over^i$ times the Poisson bracket (of order $k^{-1}$)
give relations (6.\ref{latcurr}). The conformal symmetry is realized
on the level of lattice Kac-Moody algebras by ``block spin''
homomorphisms $\CD$ corresponding to increasing maps $D:\NZ\s\rightarrow\s\NZ$
satisfying $D(n+N')=D(n)+N$\s:
\qq
\CK_{N'}\s\ni\s \J(n)\ \s\smash{\mathop{\longmapsto}
\limits^{\CD}}\s\ \J(D(n+1)-1)\s\cdots\s
\J(D(n)+1)\s \J(D(n))\s\in\s\CK_N
\qqq
\no(Recall that ${\rm Diff}_+S^1$ acts by automorphisms on continuum
Kac-Moody algebras).
\vs 1cm

\no{\bf 8.\ \ Free field representations of $\CK_N$}
\vs 0.4cm
\addtocounter{equation}{-6}
\renewcommand{\theequation}{\arabic{equation}}

As in continuum, the lattice Kac-Moody algebra may be expressed by
free fields or rather their simple deformations. The fields in
question are $\Theta(n)$ (invertible) and $\B(n),\s\Gamma(n)$ which
should be thought of as $\epsilon$-lattice versions of fields
$\ee^{i\epsilon\p\phi(\epsilon n)}$\s, $\s\beta(\epsilon n)$
\s and \s$\gamma(\epsilon n)$. Their defining relations are:
\qq
\hbox to 3cm{$\Theta(n)\s\Theta(n+1)$\hfill}&=&
q^{1/2}\s\Theta(n+1)\s\Theta(n)\s,\cr
\hbox to 3cm{$q\s\s\B(n)\s \Gamma(n)$\hfill}&=&
q^{-1}\Gamma(n)\s\B(n)+
q-q^{-1}\s,\cr
\hbox to 3cm{$\Theta(n)\s\B(n)$\hfill}&=&
q\s\B(n)\s\Theta(n)\s,\cr
\hbox to 3cm{$\Theta(n)\s\B(n+1)$\hfill}&=&
q^{-1}\B(n+1)\s\Theta(n)\s,\clabel{FrF}\cr
\hbox to 3cm{$\Theta(n)\s\Gamma(n)$\hfill}&=&
q^{-1}\Gamma(n)\s\Theta(n)\s,\cr
\hbox to 3cm{$\Theta(n)\s\Gamma(n+1)$\hfill}&=&
q\s\Gamma(n+1)\s\Theta(n)\s.
\qqq
\no All other commutators are trivial. The first relation is a lattice
version of the $u(1)$ current algebra. Second equation is a
deformation of the local oscillator algebra. The rest introduces
(essentially for convenience) a twist in the tensor product
of the two algebras.
\vs 0.3cm

Let us assume that $k$ is integer and consequently $q^{2p}=1$
where $p\equiv k+2$\s. Let us also take, for convenience,
$N$ odd. Deformed free fields $\Theta,\ \B$ and $\Gamma$
may be realized in finite-dimensional space $\CH$ with the bases
$|\un{\alpha}\rangle\otimes|\un s\rangle$ where $\un{\alpha}=(\alpha_0,
\alpha_2,\s\dots\s,\alpha_{N-1})$, $\alpha_m\in\NZ_{4p}$,
$\sum\alpha_{2m}=0$,
and $\un{s}=(s_0,s_1,\s\dots\s,s_{N-1})\s,$ \s$s_n=0,1,\s\dots\s,p-1\s$. \s
The even $\Theta(n)$ (except $\Theta(N-1)$\s) \s act diagonally,
the odd ones by shifts of $\alpha$'s.
$\B(n)$'s are the lowering and $\Gamma(n)$'s the raising operators
for the parafermionic occupation numbers $s_n<p$. More exactly, the action
of field operators is given by the formulae:
\be
\hbox to 3.5cm{$\Theta(2m)\s|\un{\alpha}\rangle\otimes|\un s\rangle$\hfill}&=&
z_{2m}\s q^{s_{2m+1}-s_{2m}+\alpha_{2m}/2}
\s|\un{\alpha}\rangle\otimes|\un s\rangle\s\ \ \ {\rm for}\ \s 2m\not=N-1,\cr
\hbox to 3.5cm{$\Theta(N-1)\s|\un{\alpha}\rangle\otimes|\un s\rangle$\hfill}&=&
z_{N-1}\s q^{s_0-s_{N-1}+\alpha_{N-1}/2}\s|\un{\alpha'}\rangle
\otimes|\un s\rangle\s,\cr
\hbox to 3.5cm{$\Theta(2m+1)\s|\un{\alpha}\rangle\otimes|\un
s\rangle$\hfill}&=&
z_{2m+1}\s q^{s_{2m+2}-s_{2m+1}}\s|\un{\alpha''}\rangle\otimes|\un s\rangle\s,
\clabel{FRFR}\cr
\hbox to 3.5cm{$\B(n)\s|\un{\alpha}\rangle\otimes|\un s\rangle$\hfill}&=&
(1-q^{-2s_n})\s|\un{\alpha}\rangle\otimes|\un{s'}\rangle\s,\cr
\hbox to 3.5cm{$\Gamma(n)\s|\un{\alpha}\rangle\otimes|\un{s}\rangle$\hfill}&=&
|\un{\alpha}\rangle\otimes|\un{s''}\rangle\ \ \ {\rm or}\ 0\ \ {\rm if}\
s_n=p-1\s.
\qqq
\no where
\qq
\un{\alpha'}&=&(\alpha_0-1,\alpha_2,\s\dots\s,\alpha_{N-3},\alpha_{N-1}+1)\s,\cr
\un{\alpha''}&=&(\alpha_0,\s\dots\s,\alpha_{2m-2},
\alpha_{2m}+1,\alpha_{2m+2}-1,\alpha_{2m+4},\s\dots\s,\alpha_{N-1})\s,\cr
\un{s'}&=&(s_0,\s\dots\s,s_{n-1},s_n-1,s_{n+1},\s\dots\s,s_{N-1})\s,\cr
\un{s''}&=&(s_0,\s\dots\s,s_{n-1},s_n+1,s_{n+1},\s\dots\s,s_{N-1})\s.
\nonumber\qqq
\no $z_n$ are arbitrary non-zero complex numbers. The above
formulae give irreducible representations of the free field algebra
$\CF_N$. They are characterized by the values of the Casimirs
$\s\Theta(n)^{4p}=z_n^{4p}\s,\s$ and
$\s\Pi\equiv q^{-1/2}\s\Theta(N-1)\s\cdots\s\Theta(1)\s\Theta(0)$
$=q^{-(N+1)/4}\prod\limits_nz_n\s$.
\vs 0.3 cm

Now, the crucial fact is that there exists a homomorphism of the
lattice Kac-Moody algebra $\CK_N$
into $\CF_N$ given by the following realization of the generators
$\J(n)$ via free fields $\Theta,\ \B$ and $\Gamma$:
\qq
\hbox to 15.2cm{$\J(n)\s\ =$\hfill}\cr
\hs*{-0.4cm}\label{FFR}\cr
\pmatrix{\hs{-0.1cm}\Theta(n)+q^{-1/2}\Theta(n)^{-1}\B(n+1)\Gamma(n)
& -\Theta(n)\B(n)+q^{-1/2}\Theta(n)^{-1}\B(n+1)
(1-\Gamma(n)\B(n))\hs{-0.06 cm}\cr
\hs{-0.1 cm}q^{1/2}\Theta(n)^{-1}\Gamma(n)& q^{1/2}\Theta(n)^{-1}
(1-\Gamma(n)\B(n))\hs{-0.06 cm}}.
\qqq
\vs 0.3 cm
\no Expressions (\ref{FFR}) are the lattice version of
the Wakimoto realization of Kac-Moody currents (5.\ref{cur}).
Due to homomorphism (\ref{FFR}), each representation of
algebra $\CF_N$ induces a representation
of $\CK_N$. In particular this holds for irreducible representations
of $\CF_N$ in space $\CH$
described above so that $\CH$ becomes a $\CK_N$-module.
The lattice Wakimoto modules obtained this way are irreducible if
$\Pi\not=q^r$ for integer $r$ (the generic case). For $\Pi=q^r$ we shall
study their reducibility mimicking the cohomological constructions
of \cite{28} which extended the original work of Felder \cite{29}
on free field
representations of the Virasoro algebra to the case of $\hat{L\s sl}(2)$
Kac-Moody algebra.
\vs 1cm

\no{\bf 9.\ \ Lattice Bernard-Felder cohomology}
\vs 0.4cm
\addtocounter{equation}{-3}
\renewcommand{\theequation}{\arabic{equation}}

We shall need an extension of free field algebra $\CF_N$
obtained by adding the invertible generator of zero mode
$\Psi(0)$ with the only nontrivial commutators
\qq
\hbox to 2.6cm{$\Theta(0)\s\Psi(0)$\hfill}&=&q^{-1/2}\s\Psi(0)\s\Theta(0)\s,\cr
\hbox to 2.6cm{$\Theta(N-1)\s\Psi(0)$\hfill}&=&q^{-1/2}\s\Psi(0)\s\Theta(N-1)
\s,\cr
\hbox to 2.6cm{$\B(0)\s\Psi(0)$\hfill}&=&q\s\Psi(0)\s\B(0)\s,\cr
\hbox to 2.6cm{$\Gamma(0)\s\Psi(0)$\hfill}&=&q^{-1}\s\Psi(0)\s\Gamma(0)\s.
\nonumber
\qqq
\no With the zero mode added, we may construct the lattice free field vertex
operator corresponding to continuum $\ee^{i\phi(\epsilon n)}$\s:
\qq
\Psi(n)&=&\Theta(n-1)\s...\s\Theta(1)\s\Theta(0)\s\Psi(0)\ \ \ \ \ \ \ \ \
\ \ {\rm for}\ n>0\s,\cr
\Psi(n)&=&\Theta(n)^{-1}...\s\Theta(-2)^{-1}
\Theta(-1)^{-1}\Psi(0)\ \ \ \ {\rm for}
\ n<0\s.\label{fermi}
\qqq
\no Notice that
\qq
\Psi(n+N)&=&q^{1/2}\s\Pi\s\s\Psi(n)\s\s=\s\s q^{-1/2}\s\Psi(n)\s\s\Pi\s.
\nonumber
\qqq
\vs 0.2cm

Let us group the lattice Wakimoto modules into families of $2p$ elements
with $\s\Theta(n)^{4p}\s$ fixed and $\Pi=q^r Z$ with $Z$ fixed and
$r=-p+1,-p+2,\s\dots\s,p$\s.
Within a fixed family we shall label the modules as $\CH_{q^r}$ according
to the value of $\Pi$.
$\Psi(0)$ may be implemented (uniquely up to normalization)
as an operator from $\CH_{q^r}$ to $\CH_{q^{r-1}}$. We shall introduce
the lattice version of the integral of the screening charge (compare
(5.\ref{SC})\s)\s:
\be
Q(n)=\Pi^{-1}\sum\limits_{m=n}^{n+N-1}\Gamma(n)\s\Psi(m)^2\s.
\qqq
\no $Q(n)$ is nilpotent in the free field representations
we consider:
\be
Q(n)^p=0
\qqq
\no (this essentially follows from the same property of $\Gamma(m)$).
As a result, we have the following complexes of vector spaces for
$r=0,1,\s\dots\s,p$\s:
\qq
&0\ \rightarrow\ \CH_{q^{-r}}\ \smash{\mathop{\rightarrow}\limits^{Q^{p-r}}}
\ \CH_{q^{r}}\ \smash{\mathop{\rightarrow}\limits^{Q^{r}}}\ \CH_{q^{-r}}\
\rightarrow\ 0&\ \s\label{short1}
\qqq
\vs -0.4cm
\qq
&0\ \rightarrow\ \CH_{q^r}\ \smash{\mathop{\rightarrow}\limits^{Q^{r}}}
\ \CH_{q^{-r}}\ \smash{\mathop{\rightarrow}\limits^{Q^{p-r}}}\ \CH_{q^r}\
\rightarrow\ 0&\ .\label{short2}
\qqq
\vs 0.2 cm
\no The important point is that for the non-generic case when $Z=1$ (i.e.
$\s\Pi$ is a $2p$-th root of unity),
the maps of the complexes are
independent of the argument $n$ of $Q$ and commute with currents $L(n)$
so that we obtain complexes of $\CK_N$-modules!
These complexes have no middle cohomology. We conjecture that the remaining
cohomology
\qq
\CH_{q^{p-r}}\supset
{\rm ker}\s\s Q^{p-r}\s\cong\s\CH_{q^{r}}\hs{-0.05cm}/\s\s{\rm im}\s\s Q^{p-r}
\qqq
\vs -0.5cm
\no and
\vs -0.7 cm
\qq
\CH_{q^r}\supset
{\rm ker}\s\s Q^r\s\cong\s\CH_{q^{-r}}\hs{-0.05cm}/\s\s{\rm im}\s\s Q^{r}
\qqq
\no gives irreducible representations of $\CK_N$.
\vs 1cm

\no{\bf 10.\ \ Vertex operators}
\vs 0.4cm
\addtocounter{equation}{-7}
\renewcommand{\theequation}{\arabic{equation}}

Up to now, we have considered only lattice currents. We shall also need
lattice WZW vertex operators corresponding to classical continuum fields
$\s h_L(x)\s$ on model space $M_{\hat{LG}}$ of $\hat{L\s sl}(2)$ Kac-Moody
group. We put (compare (5.\ref{vert})\s)
\qq
h_L(n)=\pmatrix{(\s\Pi-\Pi^{-1})\s\Psi(n)+\B(n)\s\Psi(n)^{-1}\s Q(n)&
\ \B(n)\s\Psi(n)^{-1}\cr \Psi(n)^{-1}\s Q(n)&\ \Psi(n)^{-1}}\s.\label{field}
\qqq
\no (we have omitted matrix $\pmatrix{(\Pi-\Pi^{-1})^{-1}&0\cr
0&1}$ on the right hand side of (5.\ref{vert}) since it becomes singular
for $\Pi=\pm 1$).
Two raws of $h_L(n)$ correspond to the values $\pm 1/2$ of the magnetic number
of the WZW vertex operator in spin $1/2$ representation. The vertex
operators of higher spins are polynomials of those in fundamental
representation.
\no $h_L(n)$ have twisted periodicity
\qq
h_L(n+N)=q^{-1/2}\s\s h_L(n)\pmatrix{\Pi&0\cr 0&\Pi^{-1}}\s.\label{mon}
\qqq
\no They are related to the currents by
\qq
h_L(n+1)=\J(n)\s h_L(n)
\qqq
\no and have with them the following commutation relations:
\qq
\hbox to 3cm{$\J(n)_1\s\s R^-\s h_L(n)_2$\hfill}
&=&h_L(n)_2\s\s \J(n)_1\s,\cr
\hbox to 3cm{$\J(n)_1\s\s h_L(n+1)_2$\hfill}
&=&R^+\s h_L(n+1)_2\s\s \J(n)_1\s,
\clabel{comfields}\cr
\hbox to 3cm{$\J(n)_1\s h_L(m)_2$\hfill}&=&
h_L(m)_2\s \J(n)_1\s\ \ \ {\rm for}\ m\not=n,n+1\s.
\qqq
\no Clearly, $h_L(n)$'s act in $\bigoplus\limits_{r=-p+1}^{p}{\CH_{q^r}}$.
For the generic case when $\Pi$ is not a $2p$-th root of unity,
the commutation relations of $h_L(n)$'s are given by
\qq
h_L(n)_1\s h_L(m)_2&=&h_L(m)_2\s h_L(n)_1\s D^\pm(\Pi)\s\ \ \ \ \
{\rm for}\s\ n{_>\atop^<}m\s,\ \ |n-m|<N\s,\cr
h_L(n)_1\s h_L(n)_2\s&=&R^\pm\s h_L(n)_2\s h_L(n)_1\s D^\mp(\Pi)
\label{exrel1}
\qqq
\no where the braiding matrices
\vs -0.2cm
\qq
&&\ D^\pm(\Pi)\s=\cr
&&q^{\mp 1/2}\s\pmatrix{1&0&0&\ 0\cr 0&\ 1\pm(q^{\pm 2}-1)\Pi^{\pm 1}
(\Pi-\Pi^{-1})^{-1}&\ \pm(q^{\pm 2}-1)\Pi^{\mp 1}(\Pi-\Pi^{-1})^{-1}&\ 0\cr
0&\ \mp(q^{\pm 2}-1)\Pi^{\pm 1}(\Pi-\Pi^{-1})^{-1}&\ 1\mp(q^{\pm 2}-1)\Pi^{\mp
1}
(\Pi-\Pi^{-1})^{-1}&\ 0\cr 0&0&0&\ 1}
\qqq
\no and coincide with special $6j$-symbols of quantum group
\s$\CU_q(sl(2))$ \cite{30}.
\vs 0.2cm

For the non-generic case $Z=1$, the raws of $h_L(n)$ induce the
following maps between exact sequences of vector spaces:
\qq
\matrix{\cr
\cdots&\smash{\mathop{\longrightarrow}\limits^{Q^{r}}}&\CH_{q^{-r}}&
\smash{\mathop{\longrightarrow}\limits^{Q^{p-r}}}&\CH_{q^r}&
\smash{\mathop{\longrightarrow}\limits^{Q^{r}}}&\CH_{q^{-r}}&
\smash{\mathop{\longrightarrow}\limits^{Q^{p-r}}}&\cdots\cr
\cr
&&\ \s\s\Big\downarrow h_L(n)_{a1}&&\ \s\s\s\s
\Big\downarrow h_L(n)_{a2}&&\
\s\s\Big\downarrow h_L(n)_{a1}&&&\cr
\cr
\cdots&\smash{\mathop{\longrightarrow}\limits^{Q^{r+1}}}&\CH_{q^{-r-1}}&
\smash{\mathop{\longrightarrow}\limits^{Q^{p-r-1}}}&\CH_{q^{r+1}}&
\smash{\mathop{\longrightarrow}\limits^{Q^{r+1}}}&\CH_{q^{-r-1}}&
\smash{\mathop{\longrightarrow}\limits^{Q^{p-r-1}}}&\cdots}\label{homot+}
\en
\no for $\s r=0,1,\s\dots\s,p-1\s$ and
\qq
\matrix{\cr
\cdots&\smash{\mathop{\longrightarrow}\limits^{Q^{r}}}&\CH_{q^{-r}}&
\smash{\mathop{\longrightarrow}\limits^{Q^{p-r}}}&\CH_{q^r}&
\smash{\mathop{\longrightarrow}\limits^{Q^{r}}}&\CH_{q^{-r}}&
\smash{\mathop{\longrightarrow}\limits^{Q^{p-r}}}&\cdots\cr
\cr
&&\ \s\s\Big\downarrow h_L(n)_{a2}&&\ \s\s\s\s
\Big\downarrow h_L(n)_{a1}&&\
\s\s\Big\downarrow h_L(n)_{a2}&&&\cr
\cr
\cdots&\smash{\mathop{\longrightarrow}\limits^{Q^{r-1}}}&\CH_{q^{-r+1}}&
\smash{\mathop{\longrightarrow}\limits^{Q^{p-r+1}}}&\CH_{q^{r-1}}&
\smash{\mathop{\longrightarrow}\limits^{Q^{r-1}}}&\CH_{q^{-r+1}}&
\smash{\mathop{\longrightarrow}\limits^{Q^{p-r+1}}}&\cdots}\label{homot-}
\en
\no for $\s r=1,2,\s\dots\s,p\s$. All squares in the above diagrams
are commutative! \s Define for $\s r=0,1,\s\dots\s,p\s$
\qq
\CH'_{q^r}\s\s\s\s&\s\hspace{-.5cm}
\equiv{\rm ker}\s Q^r\subset\CH_{q^r}\s,\hspace{2cm}
\CH'_{q^{-r}}\s&\hspace{-.3cm}\equiv\CH_{q^{-r}}/{\rm im\s}Q^r\s,\cr
\CH''_{q^r}\s\s\s\s&\hspace{-.5cm}
\equiv\CH_{q^r}/{\rm im\s}Q^{p-r}\s,\hspace{2cm}
\CH''_{q^{-r}}&\hspace{-.3cm}\equiv{\rm ker\s}Q^{p-r}\subset\CH_{q^{-r}}
\qqq
\no(notice the consistency of the definitions). These are the
conjectured irreducible $\CK_N$-modules at
$\Pi$ a $2p$-th root of unity.  Notice that
\s$\CH'_{q^0}=\CH''_{q^p}=\{0\}$\s.
\s Tracing diagrams (\ref{homot+})
and (\ref{homot-}) one may easily see that fields $h_L(n)$
act in spaces
\qq
\bigoplus\limits_{r=-p+1}^{p}\CH'_{q^r}\hspace{2.5cm}{\rm and}\hspace{2.5cm}
\bigoplus\limits_{r=-p+1}^{p}\CH''_{q^r}\ .
\qqq
\no These (isomorphic) spaces quantize the lattice model space
$M_{\CK_N}$.
\vs 1cm

\no{\bf 11.\ \ Free field approach to \s$\CU_q(sl(2))$}
\vs 0.4cm
\addtocounter{equation}{-10}
\renewcommand{\theequation}{\arabic{equation}}

Lattice Kac-Moody algebra $\CK_N$ is a close relative of the quantum group
$\CU_q(sl(2))$ with generators $q^{\pm 2S^3}$, $S^\pm$ and relations
\qq
q^{2S^3}\s S^\pm\s q^{-2S^3}=q^{\pm 2}\s S^\pm\s,\ \ \ \ \ \ \ \
[S^+,S^-]=(q^{2S^3}-q^{-2S^3})/(q-q^{-1})\ .\label{QGREL}
\qqq
\no Semi-classically, it was a deformation of the $N$-th power of the
Poisson-Lie
group $SL(2)^*_r$ and $SL(2)^*_r$ is a semi-classical version of
\s$\CU_q(sl(2))$.
In fact, $\CK_N$ for $N$ changing from $1$ to $\infty$ interpolates between
the quantum group $\CU_q(sl(2))$ and the continuum Kac-Moody algebra
$\hat{L\s sl}(2)$ \cite{16}.
It is then no surprise that we may realize representation
theory of $\CU_q(sl(2))$ by deformed oscillators
\qq
q\s\B\s\Gamma-q^{-1}\s\Gamma\s\B=q-q^{-1}\s.
\qqq
\no (this is in fact nothing else that the realization in terms
difference operators \cite{31}). All the algebraic constructions of
Sections 7 to 9 may be repeated in this simplified context which may
serve as a good introduction to the free field realization of $\CK_N$.
Matrix
\qq
\gamma=q\s\pmatrix{\Pi\s(1-\B\Gamma)&-(\Pi-\Pi^{-1})\s\B+\Pi\s\B\s\Gamma\B\cr
-\Pi\s\Gamma&\Pi^{-1}+\Pi\s\Gamma\s\B}\s,\label{QGCURRENT}
\qqq
\no where $\Pi,\ \Pi^{-1}$ commute with $\B$ and $\Gamma$,
satisfies the quantized version of the relations (2.21):
\qq
\gamma_1\s R^+\s\gamma_2\s(R^-)^{-1}=R^+\s\gamma_2\s(R^-)^{-1}\s\gamma_1\s.
\label{QG}
\qqq
\no These in fact are nothing else but relations (\ref{QGREL})
in disguise as may be seen by identifying $\s\gamma\s$ with the matrix
$$\pmatrix{q^{2S^3}&\ q(q-q^{-1})S^+\cr (q-q^{-1})S^-q^{2S^3}&
\ q(q-q^{-1})^2S^-S^++q^{-2S^3}}\s.$$
\no $\B$ and $\Gamma$ may be represented in $p$-dimensional space
$\CV$ with bases
$\s|\sigma\rangle\s,\ \s\sigma=0,1,\s\dots\s,p-1$\s, \s
as parafermionic annihilation and creation operators by
\qq
\B\s|\sigma\rangle=(1-q^{-2\sigma})\s|\sigma-1\rangle\s,\ \ \ \ \ \ \
\Gamma\s|\sigma\rangle=|\sigma+1\rangle\ \ \ {\rm or}\ \ 0\ \ {\rm if}\ \
\sigma=p-1\s.
\qqq
\no Via (\ref{QGCURRENT}), we then obtain a p-dimensional highest-lowest
weight $\CU_q(sl(2))$-module. $\Pi$ acts in it as multiplication by
a non-zero number. If $\Pi$ is not a $2p$-th root
of unity (the generic situation) then the module is irreducible.
In general, it is only indecomposable.
\vs 0.3cm

We shall again group the \s$\CU_q(sl(2))$-modules into
families of $2p$ elements
$\CV_{q^r}$ with $\Pi=q^rZ$, \s$r=-p+1,-p+2,\s\dots\s,p\s$.
Let us also introduce additional generator $\psi$ s.t.
\qq
\psi\s\Pi=q\s\Pi\s\psi\s,\ \ \ \psi\s\B=q\s\B\s\psi\s,\ \ \ \
\psi\s\Gamma=q^{-1}\s\Gamma\s\psi\s
\qqq
\no and $\CQ\equiv\Pi\s\Gamma\s\psi^{-2}$\s.
$\s\psi\s$ and $\s\CQ\s$ may be implemented in
$\bigoplus\limits_{r=-p+1}^{p}\CV_{q^r}$ and then $\CQ^p=0$.
In the non-generic case when $Z=1$, we obtain complexes
of \s$\CU_q(sl(2))$-modules
\qq
&0\ \rightarrow\ \CV_{q^{-r}}\ \smash{\mathop{\rightarrow}\limits^{\CQ^{r}}}
\ \CV_{q^{r}}\ \smash{\mathop{\rightarrow}\limits^{\CQ^{p-r}}}\ \CV_{q^{-r}}\
\rightarrow\ 0&\ \s\label{sh1}
\qqq
\vs -0.6cm
\qq
&0\ \rightarrow\ \CV_{q^r}\ \smash{\mathop{\rightarrow}\limits^{\CQ^{p-r}}}
\ \CV_{q^{-r}}\ \smash{\mathop{\rightarrow}\limits^{\CQ^{r}}}\ \CV_{q^r}\
\rightarrow\ 0&\ \s\label{sh2}
\qqq
\no exact in the middle. Let us denote the remaining cohomology by
\qq
\CV'_{q^r}&\equiv&{\rm ker}\s \CQ^{p-r}\subset\CV_{q^r}\s,\hspace{2.05cm}
\CV'_{q^{-r}}\s\ \equiv\ \s\CV_{q^{-r}}/{\rm im\s}\CQ^{p-r}\s,\cr
\CV''_{q^r}&\equiv&\CV_{q^r}/{\rm im\s}\CQ^{r}\s,\hspace{2.9cm}
\CV''_{q^{-r}}\s\ \equiv\s\ {\rm ker\s}\CQ^{r}\subset\CV_{q^{-r}}
\qqq
\no for $\s r=0,1,\s\dots\s,p\s$. Notice that
\s$\CV'_{q^p}=\CV''_{q^{0}}=\{0\}\s$.
\s$\CV'_{q^r}\cong\CV'_{q^r}\s$ give the irreducible highest weight
representations of \s$\CU_q(sl(2))$ with spin $S=-r/2-1/2$ (i.e.
the highest weight vector is the eigenvector of $\s q^{2S^3}\s$
with eigenvalue $\s q^{2S}\s$). Similarly, $\s\CV''_{q^r}\cong
\CV''_{q^{-r}}\s$ give the irreducible
highest weight representations of \s$\CU_q(sl(2))$ of spin $r/2-1/2$.
\vs 0.2cm

$\CU_q(sl(2))$ vertex operators are
\qq
g_0=\pmatrix{-(\Pi-\Pi^{-1})\s\psi^{-1}+\B\s\psi\s\CQ&\ B\s\psi\cr
\psi\s\CQ&\ \psi}
\qqq
\no and satisfy
\qq
\gamma\s g_0=q^2\s g_0\s\pmatrix{\Pi&0\cr 0&\Pi^{-1}}\s,\label{QGcurr}
\qqq
\no compare to the semiclassical relation (4.2).
They act in $\s\bigoplus\limits_{r=-p+1}^{p}\CV_{q^r}\s$
and, for generic case, satisfy the commutation relations
\qq
(g_0)_1\s(g_0)_2=R^\pm\s(g_0)_2\s(g_0)_1\s D^\pm(\Pi)^{-1}\s.
\qqq
\no or, for inverse matrices
\qq
g_0^{-1}=\pmatrix{-\psi&\ q\s\B\s\psi\cr \psi\CQ&\ q\s(\Pi-\Pi^{-1})\s\psi^{-1}
-q\s\B\s\psi\s\CQ}\s(\Pi-\Pi^{-1})^{-1}\s,\label{inverse}
\qqq
\vs -0.2cm
\qq
(g_0^{-1})_1\s(g_0^{-1})_2=D^\pm(\Pi)^{-1}\s(g_0^{-1})_2\s(g_0^{-1})_1\s
R^\pm\s.
\label{exrel2}
\qqq
\vs -0.3cm
\no Also
\vs -0.7cm
\qq
(g_0)_1\s\gamma_2=R^+\s\gamma_2\s(R^-)^{-1}\s(g_0)_1
\qqq
\vs -0.4cm
\no or
\vs -0.7cm
\qq
(g_0^{-1})_1\s R^+\s\gamma_2=\gamma_2\s(g_0^{-1})_1\s R^-\s.\label{CGER}
\qqq
\vs 0.2cm
\no For the non-generic case $Z=1$, raws of $g_0$ define
maps between exact sequences
\qq
\matrix{\cr
\cdots&\smash{\mathop{\longrightarrow}\limits^{\CQ^{r}}}&\CV_{q^{r}}&
\smash{\mathop{\longrightarrow}\limits^{\CQ^{p-r}}}&\CV_{q^{-r}}&
\smash{\mathop{\longrightarrow}\limits^{\CQ^{r}}}&\CV_{q^{-r}}&
\smash{\mathop{\longrightarrow}\limits^{\CQ^{p-r}}}&\cdots\cr
\cr
&&\ \s\s\Big\downarrow (g_0)_{a1}&&\ \s\s\s\s
\Big\downarrow (g_0)_{a2}&&\
\s\s\Big\downarrow (g_0)_{a1}&&&\cr
\cr
\cdots&\smash{\mathop{\longrightarrow}\limits^{\CQ^{r+1}}}&\CV_{q^{r+1}}&
\smash{\mathop{\longrightarrow}\limits^{\CQ^{p-r-1}}}&\CV_{q^{-r-1}}&
\smash{\mathop{\longrightarrow}\limits^{\CQ^{r+1}}}&\CV_{q^{r+1}}&
\smash{\mathop{\longrightarrow}\limits^{\CQ^{p-r-1}}}&\cdots}\label{homo+}
\en
\no for $\s r=0,1,\s\dots\s,p-1\s$ and
\qq
\matrix{\cr
\cdots&\smash{\mathop{\longrightarrow}\limits^{\CQ^{r}}}&\CV_{q^{r}}&
\smash{\mathop{\longrightarrow}\limits^{\CQ^{p-r}}}&\CV_{q^{-r}}&
\smash{\mathop{\longrightarrow}\limits^{\CQ^{r}}}&\CV_{q^{r}}&
\smash{\mathop{\longrightarrow}\limits^{\CQ^{p-r}}}&\cdots\cr
\cr
&&\ \s\s\Big\downarrow (g_0)_{a2}&&\ \s\s\s\s
\Big\downarrow (g_0)_{a1}&&\
\s\s\Big\downarrow (g_0)_{a2}&&&\cr
\cr
\cdots&\smash{\mathop{\longrightarrow}\limits^{\CQ^{r-1}}}&\CV_{\CQ^{r-1}}&
\smash{\mathop{\longrightarrow}\limits^{\CQ^{p-r+1}}}&\CV_{\CQ^{-r+1}}&
\smash{\mathop{\longrightarrow}\limits^{\CQ^{r-1}}}&\CV_{\CQ^{r-1}}&
\smash{\mathop{\longrightarrow}\limits^{\CQ^{p-r+1}}}&\cdots}\label{homo-}
\en
\no for $\s r=1,2,\s\dots\s,p\s$ and, consequently, $g_0$ may be defined
as an operator acting in spaces
\qq
\bigoplus\limits_{r=-p+1}^{p}\CV'_{q^r}\hspace{2.5cm}{\rm and}\hspace{2.5cm}
\bigoplus\limits_{r=-p+1}^{p}\CV''_{q^r}\ .
\qqq
\no It gives there the quantum $3j$-symbols for tensor multiplication with
the fundamental representation of $\s\CU_q(sl(2))$. \s Matrix elements
of $g_0^{-1}$ act in similar spaces but with $\CV'_{q^{0}}$ and
$\CV''_{q^p}$ excluded from the direct sums (due to the factor
$(\Pi-\Pi^{-1})^{-1}$ on the right hand side of (\ref{inverse})\s).
\vs 1cm

\no{\bf 12.\ \ Quantization of lattice  chiral WZW model}
\vs 0.4cm
\addtocounter{equation}{-19}
\renewcommand{\theequation}{\arabic{equation}}

{}From the classical expression (4.1) for the vertex-IRF transformation,
we should expect that the quantum field
\qq
g_L(n)=h_L(n)\s g_0^{-1}
\qqq
\no composed of the vertex operators of lattice Kac-Moody algebra $\CK_N$
and of quantum group $\CU_q(sl(2))$ should give the lattice quantization
of the classical chiral field $g_L$ of the WZW model. Notice
that this field acts on the diagonal product space
\qq
\bigoplus\limits_{r=-p+1}^{p}\CH_{q^r}\otimes\CV_{q^r}
\qqq
\no in the generic case when $\Pi$ is not $2p$-th root of unity and
that
\qq
g_L(n+N)=q^{-3/2}\s g_L(n)\s\gamma\s,\label{ChF}
\qqq
\no see (10.\ref{mon}) and (11.\ref{QGcurr}).
Moreover relations (10.\ref{exrel1}), (11.\ref{exrel2}) and (11.\ref{CGER})
immediately give the commutation relations
\qq
\hbox to 2.7cm{$g_L(n)_1\s g_L(m)_2$\hfill}&=&
g_L(m)_2\s g_L(n)_1\s R^{\pm}\ \ \ \ \ {\rm for}
\ \ n{_>\atop^<}m\s,\ \ |n-m|<N\s,\cr
\hbox to 2.7cm{$g_L(n)_1\s g_L(n)_2$\hfill}&=&
R^+\s g_L(n)_2\s g(n)_1\s R^-\s,\cr
\hbox to 2.7cm{$g_L(n)_1\s R^+\s\gamma_2$\hfill}&=&
\gamma_2\s g_L(n)_1\s R^-\label{CHrel}
\qqq
\no which, together with (11.\ref{QG}),
quantize expressions (6.\ref{ChPB}).
In particular the (quantum) vertex-IRF transformation,
by combining the lattice WZW vertex operators $h_L(n)$ which have
quantum $6j$-symbols as braiding matrices with
quantum $3j$-symbols $g_0^{-1}\s$, \s produced chiral fields $g_L(n)$
with $R^\pm$ braiding matrices realizing on the
lattice the program of \cite{15}.
\vs 0.2cm

In the non-generic case when $\Pi=q^r$ in $\CH_{q^r}$ and
$\CV_{q^r}$\s, there is a problem in defining the right
hand side of (\ref{ChF}) because of the $\s(\Pi-\Pi)^{-1}\s$
factor on the right hand side of (11.\ref{inverse}) which
diverges when $\Pi=\pm 1$. Nevertheless, operators
$g_L(n)$ are still well defined on the diagonal product
spaces
\qq
\bigoplus\limits_{r=-p+1}^{-1}\CH'_{q^r}\otimes\CV'_{q^r}\ \cong\
\bigoplus\limits_{r=1}^{p-1}\CH'_{q^r}\otimes\CV'_{q^r}\hspace{1cm}{\rm and}
\hspace{1cm}\bigoplus
\limits_{r=-p+1}^{-1}\CH''_{q^r}\otimes\CV''_{q^r}\ \cong\ \bigoplus
\limits_{r=1}^{p-1}\CH''_{q^r}\otimes\CV''_{q^r}\label{space}
\qqq
\no since $\s\CH'_{q^0}=\CH''_{q^p}=\CV'_{q^p}=\CV''_{q^{0}}=\{0\}\s.$
$\s g_L(n)$ do not, however satisfy on these spaces the exchange relations
(\ref{CHrel})\s ! \s The non-generic case $\Pi=q^r$ corresponds to the
values of the momenta which appear in the unitary quantization
of the continuum WZW theory \cite{28}. There, one expects existence
of continuum limit
chiral fields $g_L(x)$ acting in the space
$$\bigoplus\limits_{S=1/2}^{k/2}\CH_{k,S}\otimes\CV''_{q^{2S+1}}$$
where $\CH_{k,S}$ carries the irreducible highest weight
representation
of Kac-Moody algebra $\hat{L\s sl}(2)$ of level $k$ and spin $S\s$
(the doubling of the spins in (\ref{space}) is due to the fact
that the lattice representations should give rise to either highest
or lowest weight modules under specific continuum limit process).
As noticed by numerous authors \cite{32}-\cite{34},
the continuum fields violate the exchange relation
\qq
g_L(x)_1\s g_L(y)_2=g_L(y)_2\s g_L(x)_1\s R^\pm\ \ \ \ \ {\rm for}
\ \ x{_>\atop^<}y\s,\ \ |x-y|<2\pi\s,\label{CER}
\qqq
\no which require a bigger space, containing also non-unitary
$\hat{L\s sl}(2)$-modules in a rather complicated way.
It would be nice to have a lattice realization of such an extended
space in which (\ref{CER}) would still hold in the non-generic
case. It would be also desirable to understand whether some unitarity
conditions may be used to select the lattice quantization
of the chiral WZW model.
\vs 0.3cm

In the quantized version with generic $\Pi$, the classical
Poisson-Lie symmetries
of phase space $P^N_L$ of the lattice chiral WZW model become
quantum group symmetries. On one hand side $\CK_N$ acts in the space
of states by operators $\s\J(n)=g_L(n+1)\s g_L(n)^{-1}\s$ with
the fields satisfying the following covariance relations:
\qq
\hbox to 3cm{$\J(n-1)_1\s g_L(n)_2$\hfill}&=&
R^+\s g_L(n)_2\s\J(n-1)_1\s,\cr
\hbox to 3cm{$\J(n)_1\s R^+\s g_L(n)_2$\hfill}&=&
g_L(n)_2\s \J(n)_1\s,\cr
\hbox to 3cm{$\J(n)_1\s g_L(m)_2$\hfill}&=&
g(m)_2\s\J(n)_1\ \ \ \ \ {\rm for}\ \ m\not=n-1,n\s.
\qqq
\no One may express the covariance of the fields also in the dual fashion
by saying that the map
$$g_L(n)\ \mapsto\ h(n)^{-1}\s g_L(n)$$
generates a homomorphism of the algebra of fields $g_L(n)$
to its tensor product with the
quantum group $(SL(2)_q)^N$ \cite{35} whose matrix generators satisfying
relations
$$R^\pm\s h(n)_1\s h(n)_2=h(n)_2\s h(n)_1\s R^\pm$$
quantizing Poisson brackets (2.19) and commuting for
different $n$.
\vs 0.2cm
Similarly, \s$\CU_q(sl(2))$ acts in the space of states via the action
of $\gamma$. For the generic case, the covariance
properties of the fields are expressed
by (\ref{CHrel}) or, dually, by saying that the map
$$g_L(n)\ \mapsto\ g_L(n)\s g$$
generates a homomorphism of the field algebra into its tensor product
with $SL(2)_q$.
\vs 0.2cm

The lattice conformal symmetry given by increasing maps
$\s D:\NZ\rightarrow\NZ\s$,
$\s D(n+N')=D(n)+N\s$, induces, by decimation
$$g_L(n)\s\mapsto\s g_L(D(n))\s,$$
\no a homomorphism of the field algebras implemented by maps between
the corresponding spaces of states. It remains to understand whether
some sort of lattice Sugawara construction \cite{36}
survives on the lattice (it does for the $u(1)$ case).
This is the principal open problem of the lattice
WZW theory.
\vs 0.3cm

In the above exposition, we have concentrated on the left-moving
part of the WZW theory. Right-handed piece of the lattice model
may be constructed in exactly the same way, replacing \s$q\rightarrow
q^{-1}\s$.
\s The left-right space of states
may then be obtained by combining the spaces of representations
of the left and right lattice Kac-Moody algebras as we combined
those of the left Kac-Moody algebra and those of the quantum group.
We leave the details as an exercise to the reader. The quantum
group symmetry of chiral models residing in the monodromy of the
chiral fields disappears from the real theory which continues
to possess a pair of lattice Kac-Moody symmetries.

\end{document}